 \documentclass[aps,prb,twocolumn,superscriptaddress,showpacs,amsmath,amssymb]{revtex4}

\usepackage{graphicx}
\usepackage{dcolumn} 
\usepackage{bm}      
\usepackage{verbatim}%
\usepackage{epstopdf}

\begin{document}


\title
{
Molecular Dynamics Simulation of Ga Penetration along {$\bm \Sigma 5$} Symmetric Tilt Grain Boundaries in an Al Bicrystal
}


\author{ Ho-Seok Nam }
\email
{hnam@princeton.edu}
\affiliation
{
Department of Mechanical and Aerospace Engineering, Princeton University, Princeton, New Jersey 08544, USA
}

\author{ David J. Srolovitz }
\affiliation
{
Department of Mechanical and Aerospace Engineering, Princeton University, Princeton, New Jersey 08544, USA
}
\affiliation
{
Department of Physics, Yeshiva University, New York, New York 10033, USA
}

\date{\today}

\begin{abstract}

Liquid metal embrittlement (LME) is a common feature of systems in which a low melting point liquid metal is in contact with another, higher melting point, polycrystalline metal.  While different systems exhibit different LME fracture characteristics, the penetration of nanometer-thick liquid metal films along the grain boundary is one of the hallmarks of the process.  We employ EAM potentials optimized for Al-Ga binary alloys in a series of MD simulations of an Al bicrystal (with a $\Sigma 5~ 36.9 \,^{\circ} /[010]$ symmetric tilt boundary) in contact with liquid Ga with and without an applied stress.  Our simulations clarify the mechanism of LME and how it is affected by applied stresses.  The interplay of stress and penetrating Ga atoms leads to the nucleation of a train of dislocations on the grain boundary below the liquid groove root which climbs down the grain boundary at a nearly constant rate.  The dislocation climb mechanism and the Ga penetration are coupled.  While the dislocations do relax part of the applied stress, the residual stresses keep the grain boundary open, thereby allowing more, fast Ga transport to the penetration front (i.e., Ga layer thickening process).  The coupled Ga transport and ``dislocation climb'' is the key to the anomalously fast, time-independent penetration of Ga along grain boundaries in Al.  The simulations explain a wide range of experimental observations of LME in Al-Ga the literature.

\end{abstract}

%








\maketitle


\section{\label{sec:level_1intro} Introduction}

There are many examples in which very deep grooves form at the intersections of grain boundaries and the surface in systems in which a liquid metal is in contact with a polycrystalline solid.  In some systems, such as Al-Ga, Zn-Ga, Cu-Bi and Ni-Bi, the liquid film quickly penetrates deep into the solid along the grain boundary and leads to brittle intergranular fracture under the influence of even modest stresses.  This is a form of liquid metal embrittlement (LME).  Although this phenomenon is quite common in material processing, LME is not well understood.  LME is particularly important in nuclear reactor scenarios in which liquid metals are used as coolants and as spallation targets.

The Al-Ga couple is a particularly well-known LME system.  Several studies have shown that the maximum load a polycrystalline Al sample in contact with liquid Ga can sustain decreases as the quantity of Ga on the grain boundaries increases (characterized by exposure time of these Al samples to Ga) eventually leading to intergranular brittle fracture.~\cite{Eynolds:AlGaCleavage, Chu:AlGaFracture, Uan:AlGaFracture}  Transmission electron microscopy~\cite{Hugo:AlGaTEM1998, Hugo:AlGaBicrystalTEM1999, Hugo:AlGaAtomicModelTEM2000} (TEM), scanning electron microscopy~\cite{Kozlova:AlGaSEM2005, Kozlova:AlGaSEM2006} (SEM), and  synchrotron radiation micro-radiography studies~\cite{Tsai:AlGaSRmuR, Pereiro-Lopez:AlGaPRL2005, Ludwig:AlGaBicrystal2005, Pereiro-Lopez:AlGaPolycrystal2004} all show that liquid Ga penetrates into grain boundaries in Al at a remarkable rate, leading to a distinct channel morphology.  The penetration of liquid Ga along the grain boundaries produces wetting layers with thickness ranging from several monolayers~\cite{Hugo:AlGaTEM1998, Hugo:AlGaBicrystalTEM1999, Hugo:AlGaAtomicModelTEM2000} to several hundred nanometers,~\cite{Pereiro-Lopez:AlGaPRL2005, Ludwig:AlGaBicrystal2005, Pereiro-Lopez:AlGaPolycrystal2004} even in the absence of an applied load.  Interestingly, the rate of propagation of such liquid layers is strongly influenced by even very small stresses.~\cite{Ludwig:AlGaBicrystal2005, Pereiro-Lopez:AlGaPolycrystal2004}  Based on these observations, the embrittlement of Al by liquid Ga is considered to be associated with the fast penetration of liquid gallium, resulting in rapid changes in grain boundary structure/bonding and in many properties of the polycrystal (e.g., fracture strength).~\cite{Pereiro-Lopez:AlGaPolycrystal2004}

Thermodynamically, wetting of a grain boundary by a liquid metal should be expected when the spreading coefficients $S$ satisfies: $S = \gamma_{GB} - 2 \gamma_{SL} \geq 0$, where $\gamma_{GB}$ and $\gamma_{SL}$ are the free energies of the grain boundary and solid-liquid interface, respectively.~\cite{Glickman:DCM}  However, thermodynamic arguments do not explain the liquid channel morphology, the Ga penetration kinetics, or the atomistic mechanism of Ga penetration.  The anomalously fast, time-independent penetration rate (several $\mu$m/s at room temperature~\cite{Hugo:AlGaTEM1998, Hugo:AlGaBicrystalTEM1999, Hugo:AlGaAtomicModelTEM2000, Pereiro-Lopez:AlGaPRL2005, Ludwig:AlGaBicrystal2005, Pereiro-Lopez:AlGaPolycrystal2004}) of very long nanometer-thick liquid films cannot be explained in terms of the classical Mullins grain boundary grooving~\cite{Mullins:GBGbyDv} nor by normal grain boundary diffusion.~\cite{Vogel:GBInstability1971}

Various models have been proposed to explain the kinetics and atomistic mechanisms by which the liquid phase penetrates quickly along grain boundaries, including corrosive dissolution,~\cite{Fradkov:CorrosiveDissolution} mixed diffusion-dissolution,~\cite{Bokstein:DiffusionDisolution} self indentation-internal solution,~\cite{Glickman:SIIS} coherency stresses,~\cite{Rabkin:CoherencyStresses} amorphous grain boundary/liquid transformation,~\cite{Desre:AmorphousFormation} and others.~\cite{Joseph:LMEreview}
However, even for the well-known Al-Ga LME couple case, many fundamental questions remain.  For example, the degree to which Ga penetration along grain boundaries occurs in the complete absence of internal or external stresses is still unsettled.  Furthermore, the role of stress on grain boundary penetration is not well understood.  Further issues include whether LME is essentially ``replacement-like'' (Ga atoms replace Al atoms at the grain boundary and the Al atoms are transported away) or ``invasion-like''~\cite{Pereiro-Lopez:AlGaPRL2005} (Ga atoms insert into the grain boundary without replacing Al atoms).  The latter process must generate stresses, which could be relieved by plastic deformation as the liquid layer thickens.  Clearly, the mechanisms by which grain boundary penetration occurs remain unsettled.  

The penetration of a liquid phase along grain boundaries is a complex phenomenon, involving several different types of simultaneous processes; e.g., dissolution/reprecipitation, liquid groove formation, grain boundary diffusion, and grain boundary segregation.  Because of the interplay between the underlying phenomena that occur in LME, it has been difficult to design experiments that can be easily interpreted to understand which processes control the phenomena and which are simply parasitic.  We approach LME by performing molecular dynamics (MD) simulations of an Al bicrystal in contact with liquid Ga (with and without an applied stress).  The advantage of a simulation approach is that it is much easier to interrogate both microscopic and macroscopic events that occur during LME with a resolution rarely accessible to experiments ({\AA ngstroms} and picoseconds).  As a first step, we tune a set of embedded-atom method (EAM) potentials to reproduce the experimental solid-liquid Al-Ga binary phase diagram.  We investigate how Ga penetrates along the grain boundaries, the degree to which Al dissolves into the liquid Ga, the relative displacement of the grains, the stress distribution within the solid, and physical parameters that are useful for interpreting the penetration mechanism (e.g., grain boundary diffusivity, segregation of Ga to the grain boundary, elastic constants).  Based on these simulation results, we propose a new mechanism for LME and compare it with general trends gleaned from a series of experimental studies in the literature.  A brief report on this topic, focused upon different grain boundaries, was reported in Ref.~\onlinecite{HoseokNam:LMEPRL}.

\section{\label{sec:level_2simul} Simulation Method}

\subsection{\label{sec:sublevel_21ElemPototential} Interatomic potentials for Al and Ga}

A first step in the simulation of LME in the Al-Ga system is the development of a reasonable description of the atomic interactions.  While several potentials have been developed for elemental Al (Ref.~\onlinecite{VoterChen:EAMNiAl, Foiles:EAMNiAl, Johnson:EAMAl, Hoagland:EAMAl, MeiDavenport:EAMAl, Cai:EAMfccMetals, Mishin:EAMfccMetals} and one for Ga (Ref.~\onlinecite{Baskes:MEAMGa}), we are unaware of the development of potentials that were optimized to reproduce the properties of the binary Al-Ga system.  The Al potentials listed here are of the EAM form, while the Ga potential is a modified embedded-atom method (MEAM) potential (that includes angular terms)~\cite{Baskes:MEAMGa}.  While the MEAM Ga potential reproduces many of the properties of Ga, we focus here on the computationally more efficient EAM class of potential for both Al and Ga.  To this end, we develop Al-Ga EAM potentials of the EAM type originally developed by Mei and Davenport~\cite{MeiDavenport:EAMfccMetals} for Al.  The parameters in the Mei and Davenport potentials were fitted to a series of basic  properties for which data were available, including the cohesive energy, lattice constant, and elastic constants of the face-centered cubic (fcc) crystal.  The advantage of this potential is that it is analytical and hence easily adjustable to obtain desired thermodynamic properties and extendable to binary alloys.

The total energy of the system is given by the usual EAM form:
\begin{equation}
\label{eq:EAM_TotalEnergy}
E ~ = ~ \sum_{i} \left[ F_{s_i} ( \bar \rho_i )
~ + ~ \frac {1} {2} \sum_{j \neq i} \phi_{s_i s_j}  ( r_{ij} ) \right], 
\end{equation}
where $ F_{s_i} ( \rho )$ is the energy associated with embedding atom of type $s_i$ in a uniform electron gas of density $\rho$ and $\phi_{s_i s_j} ( r )$ is a pairwise interaction between atoms of type $s_i$ and $s_j$ separated by a distance $r$.  
The pair potential term $\phi(r)$ is chosen to take the form
\begin{equation}
\label{eq:EAM_PairPotential}
\phi (r) = - \phi_{0} \left[ 1 + \delta \left( \frac{r}{r_{0}} - 1 \right) \right]
\exp \left[ - \gamma \left( \frac{r}{r_{0}} - 1 \right) \right], 
\end{equation}
where $r_{0}$ is the first-nearest-neighbor distance in the reference structure and $\phi_0$, $\gamma$ and $\delta$ are fitting parameters.
The electron density is given by
\begin{equation}
\label{eq:EAM_ElectronDensity}
\bar \rho_i = \sum_{j(\ne i)} f ( r_{ij} ), 
\end{equation}
where the atomic electron density $f(r)$ was numerically determined~\cite{MeiDavenport:EAMfccMetals} so that the electron density of each atom in the reference structure, $\rho_{\rm ref}$, satisfies the following relation: 
\begin{equation}
\label{eq:EAM_DensityFitting}
\rho_{\rm ref} = \sum_{m=1}^{3} z_{m} f(r) = \rho_{e} \exp \left[ - \beta \left( \frac{r}{r_{0}} - 1 \right) \right].  
\end{equation}
Here, $z_{m}$ is the number of $m$th-nearest-neighbor atoms in the reference structure and the parameter $\beta$ quantifies the distance over which the electron density decays away from an atom position.
Accordingly, the embedding function takes the form~\cite{MeiDavenport:EAMAl} (including interactions up to third-nearest-neighbors):
\begin{eqnarray}
\label{eq:EAM_EmbeddingEnergy}
F ( \rho ) = 
&& - E_{0} \left[ 1 - \frac{\alpha}{\beta} \ln \left( \frac{\rho}{\rho_{e}} \right) \right] 
\left( \frac{\rho}{\rho_{e}} \right) ^ { \frac{\alpha}{\beta} }
\nonumber\\
&& + \frac {1} {2} \phi_{0} \sum_{m=1}^{3} z_{m} \exp \left[ - ( \sqrt {m} - 1) \gamma \right]
\nonumber\\
&& \times \left[ 1 + ( \sqrt{m} - 1) \delta - \sqrt{m} ( \frac{\alpha}{\beta} ) \ln \left( \frac{\rho}{\rho_{e}} \right) \right]
\nonumber\\
&& \times \left( \frac{\rho}{\rho_{e}} \right) ^ { \sqrt{m} \frac{\gamma}{\beta} }
\end{eqnarray}
so that the total energy of reference system as a function of dilation  satisfies the universal binding energy relation:~\cite{Rose:EAMUniversalFeatures} 
\begin{equation}
\label{eq:EAM_UniveralRelation}
E(r) = - E_{0} \left[ 1 + \alpha \left( \frac{r}{r_{0}} - 1 \right) \right]
\exp \left[ - \alpha \left( \frac{r}{r_{0}} - 1 \right) \right], 
\end{equation}
with $\alpha = \sqrt{ 9B\Omega / E_0} $, where $E_{0}$, $B$, and $\Omega$ are the cohesive energy,  bulk modulus, and  atomic volume, respectively.
The potential interactions were truncated between the third- and forth-nearest-neighbor shells of a static fcc crystal.

For Al, we used the parameter set suggested by \citeauthor{SturgeonLaird:EAMAl}~\cite{SturgeonLaird:EAMAl} (which is a modification of those originally obtained by \citeauthor{MeiDavenport:EAMAl}~\cite{MeiDavenport:EAMAl} to reproduce the appropriate melting point of Al).  Based on the same formalism as Mei and Davenport, we fitted EAM potential parameters for Ga using data obtained from first principles with a reference fcc lattice~\cite{Baskes:MEAMGa} (see Table~\ref{tab:table1}).  Parameters $E_{0}$, $r_{0}$, and $\alpha$ were directly determined from the cohesive energy $E_{c}$, lattice parameter $a$, and bulk modulus $B$, while  $\gamma$ and $\delta$ were obtained by optimization against the other elastic constants, vacancy formation energy, and melting point.  The potential parameters for Al and Ga are shown in Table~\ref{tab:table2}.
\begin{table}[!tbp]
\caption
{\label{tab:table1}
Basic input data used in fitting the parameters for Al[\onlinecite{MeiDavenport:EAMAl}] and Ga[\onlinecite{Baskes:MEAMGa}]. The numbers in parentheses are calculated values.
}
\begin{ruledtabular}
\begin{tabular}{ccc}
properties & Al & fcc-Ga \\
\hline
$E_{c}$ (eV/atom) & 3.39 (3.39) & 2.897 (2.897) \\
$a$ (\AA)         & 4.05 (4.05) & 4.247 (4.247) \\
$B$ (GPa)         &  76   (76)  &  52    (52)   \\
\end{tabular}
\end{ruledtabular}
\end{table}

\begin{table*}[!tbp]
\caption
{\label{tab:table2}
Parameters for the Al-Ga EAM potential.  Parameters for Al were developed by \citeauthor{SturgeonLaird:EAMAl}.~\cite{SturgeonLaird:EAMAl} $E_{0}$ and $\phi_{0}$ are in units of eV, $r_{0}$ is in units of {\AA}, and the other parameters are dimensionless.
}
\begin{ruledtabular}
\begin{tabular}{cccccccc}
   & $E_{0}$ & $\phi_{0} $ & $r_{0}$ & $\alpha$ & $\beta$ & $\gamma$ & $\delta$ \\
\hline
Al & 3.39 & 0.1318 & 2.8638 & 4.60 & 7.10 & 7.34759 & 8.45 \\
Ga & 2.897 & 0.064 & 3.003  & 4.42 & 7.10 & 7.8     & 5.2  \\
Al-Ga & -  & 0.075 & 2.933  &  -   &  -   & 5.7     & 6.3  \\
\end{tabular}
\end{ruledtabular}
\end{table*}

\begin{figure}[!tbp]
\includegraphics[width=0.40\textwidth]{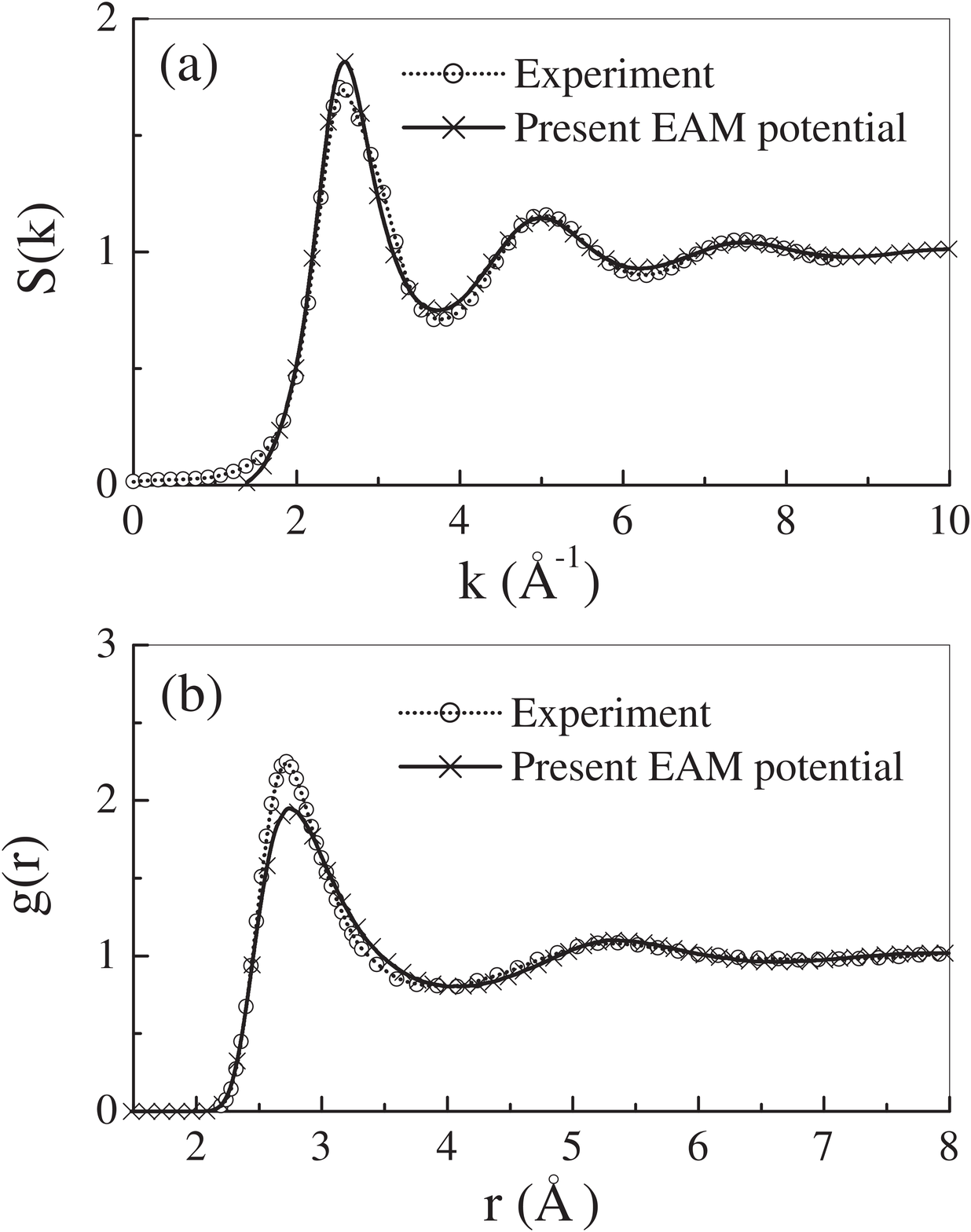}
\caption
{ \label{fig:STFnRDF}
(a) Structure factor and (b) radial distribution function for liquid Ga at 959 K for the present EAM potential and experiment.~\cite{Bellissent-Funel:GaSTF}
}
\end{figure}

\begin{figure}[!tbp]
\includegraphics[width=0.40\textwidth]{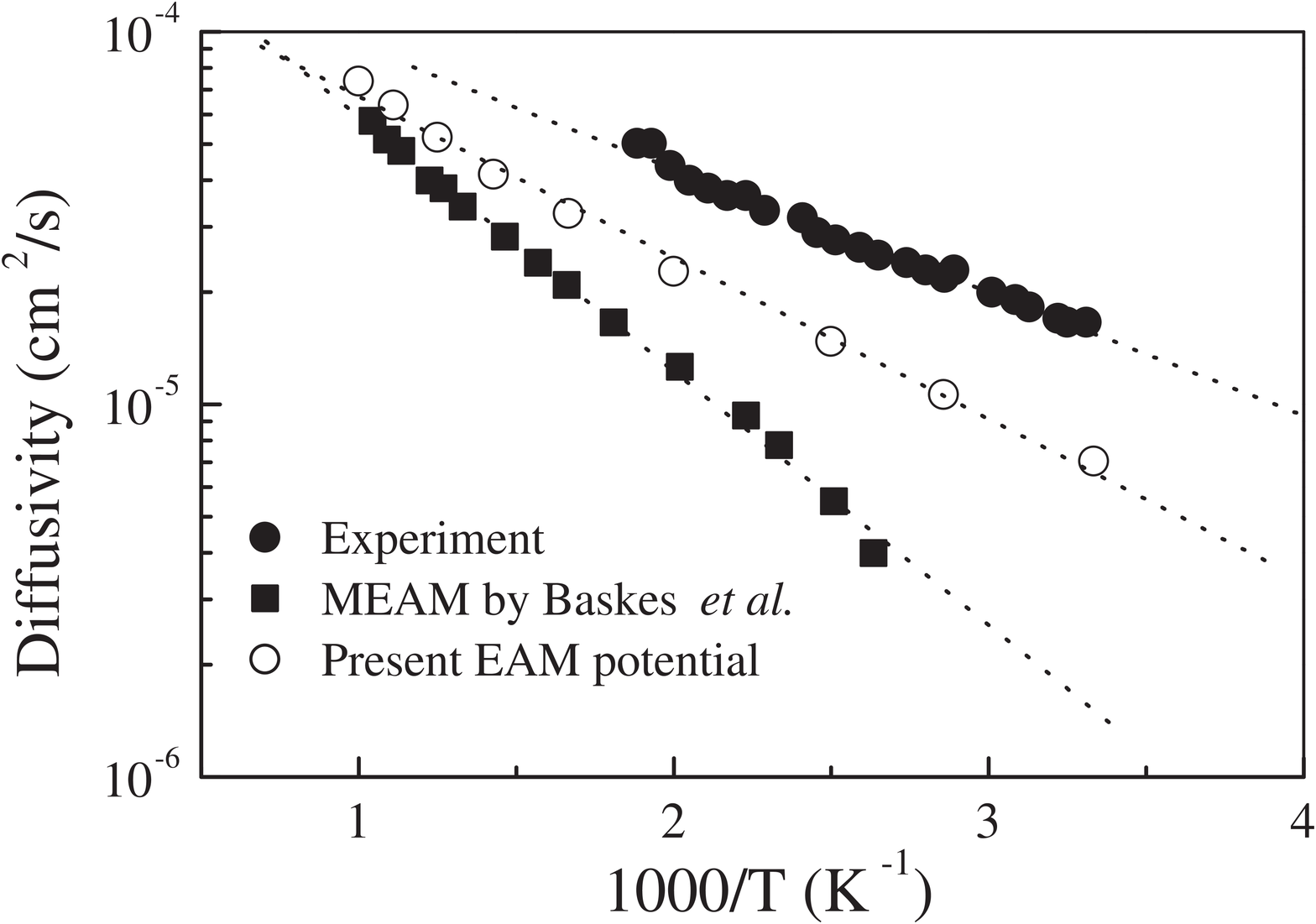}
\caption
{ \label{fig:DiffusivityGaSelf}
Arrhenius plot of the self-diffusivity of liquid Ga obtained using the present EAM potenial, the MEAM potential by Baskes {\em et al.},~\cite{Baskes:MEAMGa} and experiment.~\cite{Riedl:DiffGaSelf} 
}
\end{figure}

Unfortunately, the new potential for Ga predicts that the fcc crystal structure has a lower energy than the experimentally observed A11 $\alpha$-Ga structure.  This is not a particularly important problem, however, since in the present situation, we focus only on liquid Ga.
Fortunately, the liquid properties of our model Ga potential are in good agreement with experiment.  The melting point of fcc Ga (obtained via microcanonical ensemble molecular dynamics simulations of solid-liquid coexistence~\cite{Morris:CoexistingMD}) is 305 K which is only about $3\%$ higher than the experimental value for $\alpha$-Ga.  Figure~\ref{fig:STFnRDF} shows the structure factor $S(k)$ and the pair correlation function $g(r)$ at 959 K obtained using this potential in molecular dynamics simulation and from experiment.~\cite{Bellissent-Funel:GaSTF} Except for the first peak, the calculated $S(k)$ is in good agreement with experimental data both in peak heights and positions.  The pair correlation function shows a broad first peak centered at $\sim 2.8$ {\AA} and a very weak second peak at $\sim 5.4$ {\AA}, in good agreement with the experimental data.  However, the first peak obtained using the present potential is slightly broader than that from experiment.  This small difference is consistent with that found using the MEAM potential.~\cite{Baskes:MEAMGa}  The diffusivity of liquid Ga (Fig.~\ref{fig:DiffusivityGaSelf}) at high temperature obtained using the present potential is in better agreement with experiment~\cite{Riedl:DiffGaSelf} than that obtained using the MEAM potential.

\subsection{\label{sec:sublevel_22AlloyPotential} Potentials for the Al-Ga alloy system}

The simple analytic, EAM potential form can be easily extended to multi-component systems.  For EAM binary alloys, we must fix seven functions $f_{Al}(r)$, $f_{Ga}(r)$, $F_{Al}(r)$, $F_{Ga}(r)$, $\phi_{Al-Al}(r)$, $\phi_{Ga-Ga}(r)$ and $\phi_{Al-Ga}(r)$.  The first six of these are transferable from the two monatomic systems, Al and Ga.

\begin{figure}[!tbp]
\includegraphics[width=0.40\textwidth]{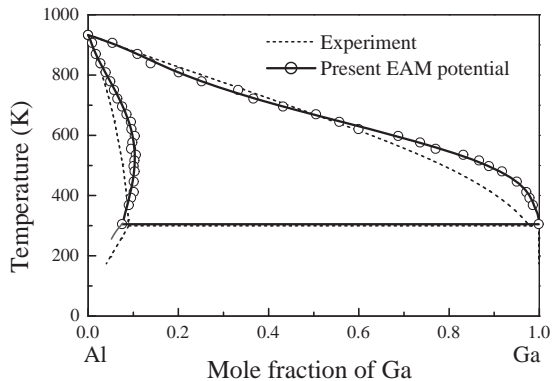}
\caption
{ \label{fig:PhaseDiagram}
Calculated and experimental~\cite{Massalski:BinaryAlloyPhaseDiagrams} phase diagram of Al-Ga binary alloy.
}
\end{figure}

We can determine the remaining function, $\phi_{Al-Ga}(r)$, by fitting parameters $\phi_{0}$, $r_{0}$, $\gamma$, and $\delta$ to alloy properties.  Here, we choose this function to reproduce the binary solid-liquid alloy phase diagram of the Al-Ga system.  Of particular importance in the LME of Al by Ga, is the solubility of Ga in Al and {\it vice versa}.  Our earlier work on general trends in binary phase diagrams as a function of atomic interactions~\cite{HoseokNam:PhaseDiagram} provides some guidance here.  
The equilibrium solid-liquid phase boundaries are monotonic functions of the fitting parameters $\phi_{0}$, $r_{0}$, $\gamma$, and $\delta$.  We initially chose $\phi_{0}$ to be the geometric mean of the monatomic pair potentials and the parameters $r_{0}$, $\gamma$, and $\delta$ to be the arithmetic mean of those of Al and Ga and then made small adjustments to them to obtain reasonable agreement with experiment.  The solid-liquid alloy phase diagram was obtained using the Gibbs-Duhem integration method, as described in Ref.~\onlinecite{Kofke:GibbsDuhemMolPhys} and \onlinecite{Hitchcock:JCP}.  The potential parameters  are given in Table~\ref{tab:table2}.  Figure~\ref{fig:PhaseDiagram} shows the calculated and experimental~\cite{Massalski:BinaryAlloyPhaseDiagrams} phase diagram for the Al-Ga alloy system.  The agreement between the two is very good except near the pure Ga end of the phase diagram(where it is still pretty good).

\begin{figure}[!tbp]
\includegraphics[width=0.40\textwidth]{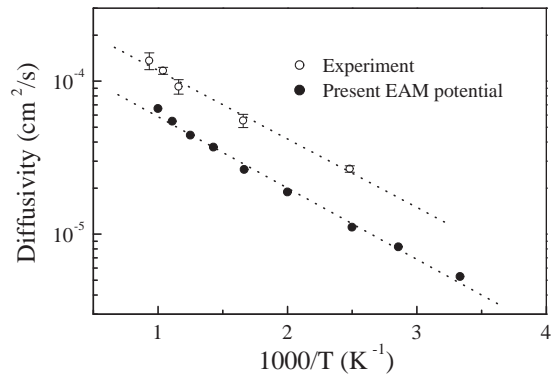}
\caption
{ \label{fig:DiffusivityAlinGa}
Arrhenius plot of Al diffusivity in molten Ga for the present EAM alloy potential and experiment.~\cite{Brauer:DiffAlinGa} 
}
\end{figure}

We also measure the diffusivity of Al in liquid Ga using the present alloy potential.  As seen in Fig.~\ref{fig:DiffusivityAlinGa}, the predicted diffusivity has the same activation energy as that found in experiment and the absolute diffusivities agree to within a factor of two.  This is considered excellent agreement for diffusivity measurements.

\subsection{\label{sec:sublevel_23LMEsimul} Molecular dynamics simulations of LME}

\begin{figure}[!tbp]
\includegraphics[width=0.48\textwidth]{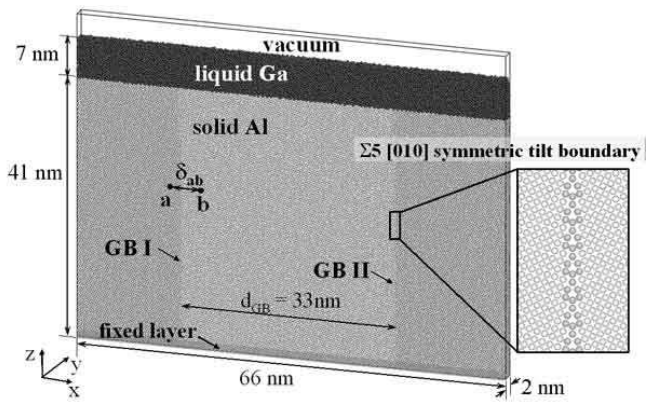}
\caption
{ \label{fig:Geometry}
Simulation cell containing two Al grains in contact with liquid Ga.  The enlarged picture shows the atomic structure of the $\Sigma 5~ 36.9 \,^{\circ} /[010]$ symmetric tilt boundary.
}
\end{figure}

All of the simulations described below were performed in a three dimensional Al bicrystal sample that is in contact with  liquid Ga.  Figure~\ref{fig:Geometry} shows a schematic illustration of the simulation cell and initial atomic configuration of the grain boundary.  We first constructed a periodic Al bicrystal by joining two unrelaxed and rotated fcc crystals and a separate liquid Ga phase (the structure was obtained by heating to above its melting temperature).  After relaxing these structures at the desired temperature by MD, we brought the two systems together into a single simulation box with a liquid layer on top of the bicrystal in such a way that the grain boundaries intersect the solid-liquid interface.  We imposed periodic boundary condition in $x$ (normal to grain boundaries) and $y$ directions, fixed one or two bottom layers of atoms in the solid to maintain the bicrystal structure and left the top surface free (i.e., there is a vacuum above the liquid).  Because of the periodic boundary conditions, two identical grain boundaries (noted as `GB I' and `GB II' in Fig.~~\ref{fig:Geometry}) are contained in a simulation cell and the solid bicrystal is effectively equivalent to a bamboo-like polycrystal with periodic grain boundaries separated by grain size $d_{GB}$ in $x$ direction.

Since the liquid Ga penetrates deep into the grain boundaries during the simulations ($\sim 70$ ns), we employ a relative thick bicrystal (at least 40 nm).  Furthermore, in order to minimize interactions between the two adjacent liquid grooves, the simulation cell dimension in the $x$ direction was also large ($\sim 66$ nm).  Finally, the simulation cell dimension in the $y$ direction was chosen to be only 5 cubic unit cell lattice parameters ($\sim 2$ nm - greater than the range of the EAM potentials).  A typical simulation cell contained $\sim 350,000$ atoms.

Our simulation study focuses primarily on the $\Sigma 5~ 36.9 \,^{\circ} /[010]$ symmetric tilt grain boundary.  We chose this special boundary as a starting point because it is particularly well-studied and has a simple structure.

\begin{figure}[!tbp]
\includegraphics[width=0.35\textwidth]{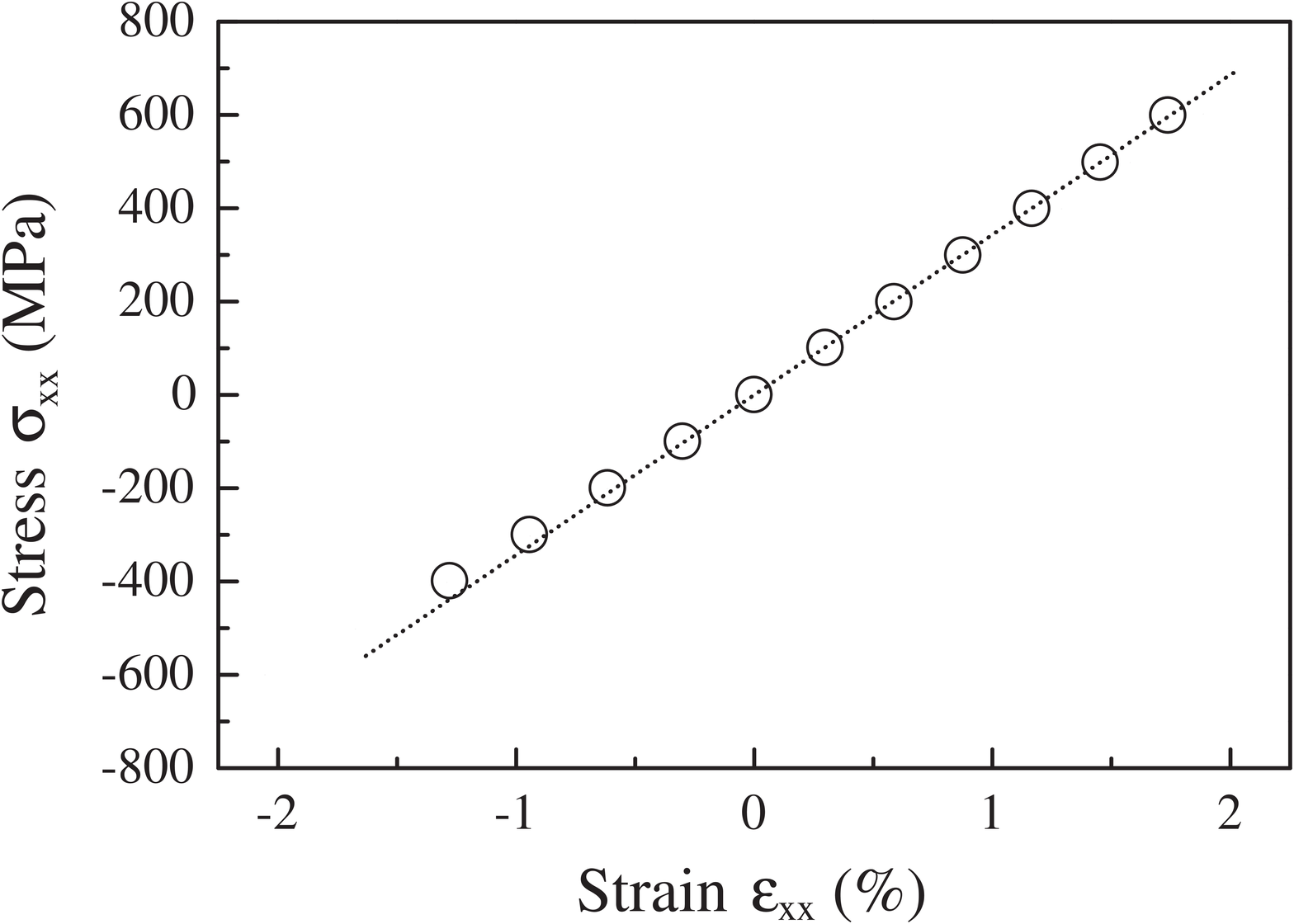}
\caption
{ \label{fig:FlowCurve}
Stress versus strain for a periodic bicrystal containing a $\Sigma 5~ 36.9 \,^{\circ} /[010]$ symmetric tilt boundary at 600 K.
}
\end{figure}

The simulations were performed under uniform, tensile loading conditions.  In a macroscopic sample, most of the load is carried by the solid, far from the surface.  However, in atomistic simulations, the sample is necessarily small and there is no material far from the surface; hence, the effect of boundary conditions for constant stress ($NPT$) and constant strain ($NVT$) simulations can be quite different.  Since the goal is to model macroscopic loading conditions, we performed the simulations under constant strain conditions, where the strain was chosen to provide the desired uniaxial stress in the bulk bicrystal.  To accomplish this, we applied a constant strain by fixing the length of the simulation cell in the $x$ and $y$ directions.  Note that the elastic modulus of solid bicrystals depends on the crystalline orientation of the two grains as well as the grain boundary characteristics.  To find the stress that corresponds to the fixed length simulation cell, we measured the stress-strain curve in a bicrystal cell that is periodic in all directions and contains no liquid.  This is shown in Fig.~\ref{fig:FlowCurve}.  For the $\Sigma 5~ 36.9 \,^{\circ} /[010]$ symmetric tilt grain boundary, the stress-strain relation of the bicrystal cells shows elastic behavior up to strain of $\sim 2 \%$.

The MD simulations were performed in the $NVT$ (canonical) ensemble using the LAMMPS (Large-scale Atomic/Molecular Massively Parallel Simulator) code.~\cite{Plimpton:ParallelAlgorithm,LAMMPS:homepage} The equations of motion were integrated using the velocity Verlet method.  The simulations were performed at $T=600$ K and with an applied uniaxial stress in the 0 to 500 MPa range.  Total simulation time was at least 70 ns ($\sim 3 \times 10^7 \Delta t$, where $\Delta t=$2.5 fs represents the time step for the integration of the equation of motion).  All simulations in this study were performed on an IBM BlueGene/L machine at Princeton University with up to 2048 processors.

\section{\label{sec:level_3MD} Penetration of Gallium into ${\bm \Sigma 5}$ symmetric tilt boundaries with and without applied stress}

\begin{figure*}[!tbp]
\includegraphics[width=0.32\textwidth]{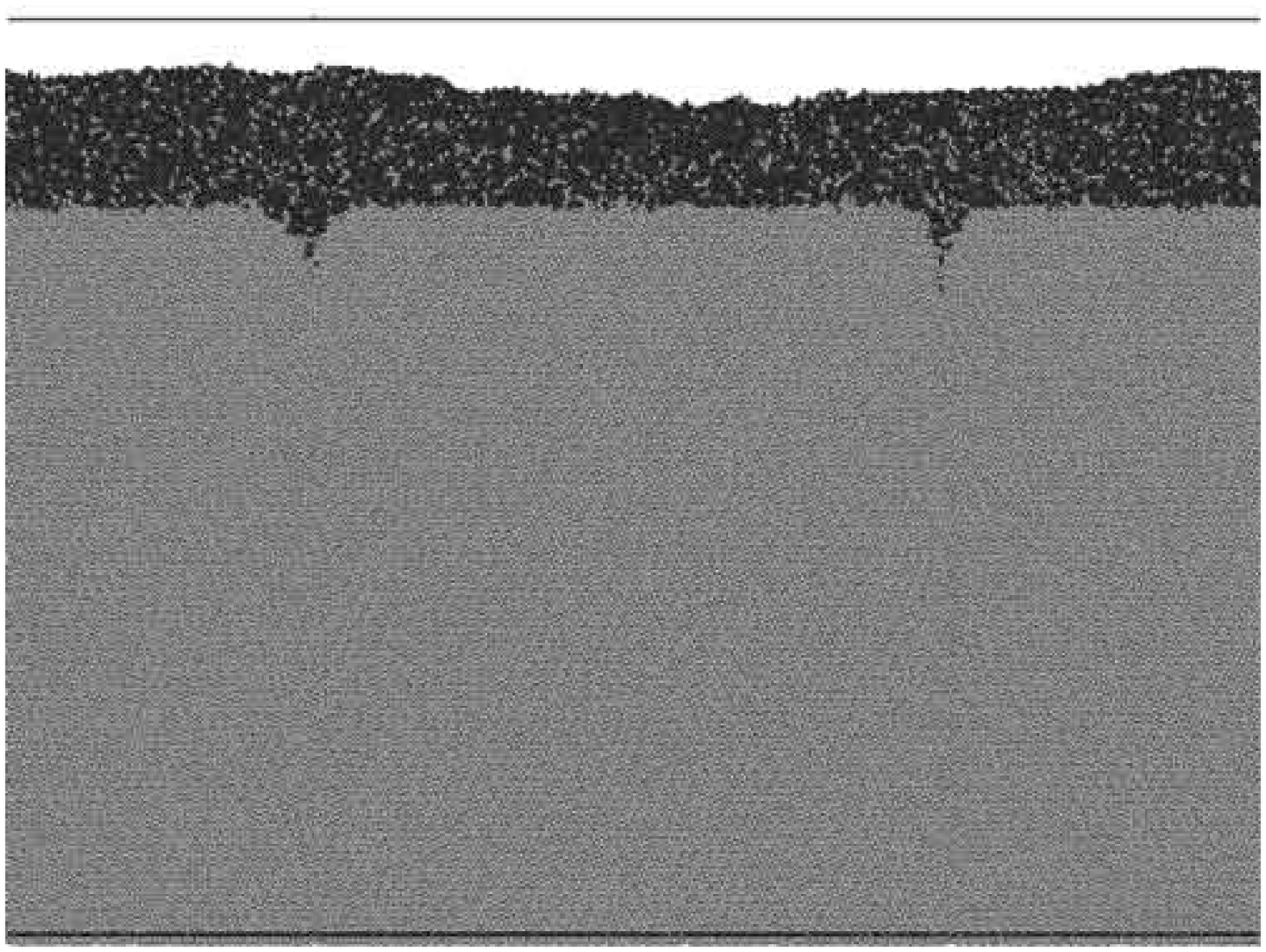}
\hspace*{0.1cm}
\includegraphics[width=0.32\textwidth]{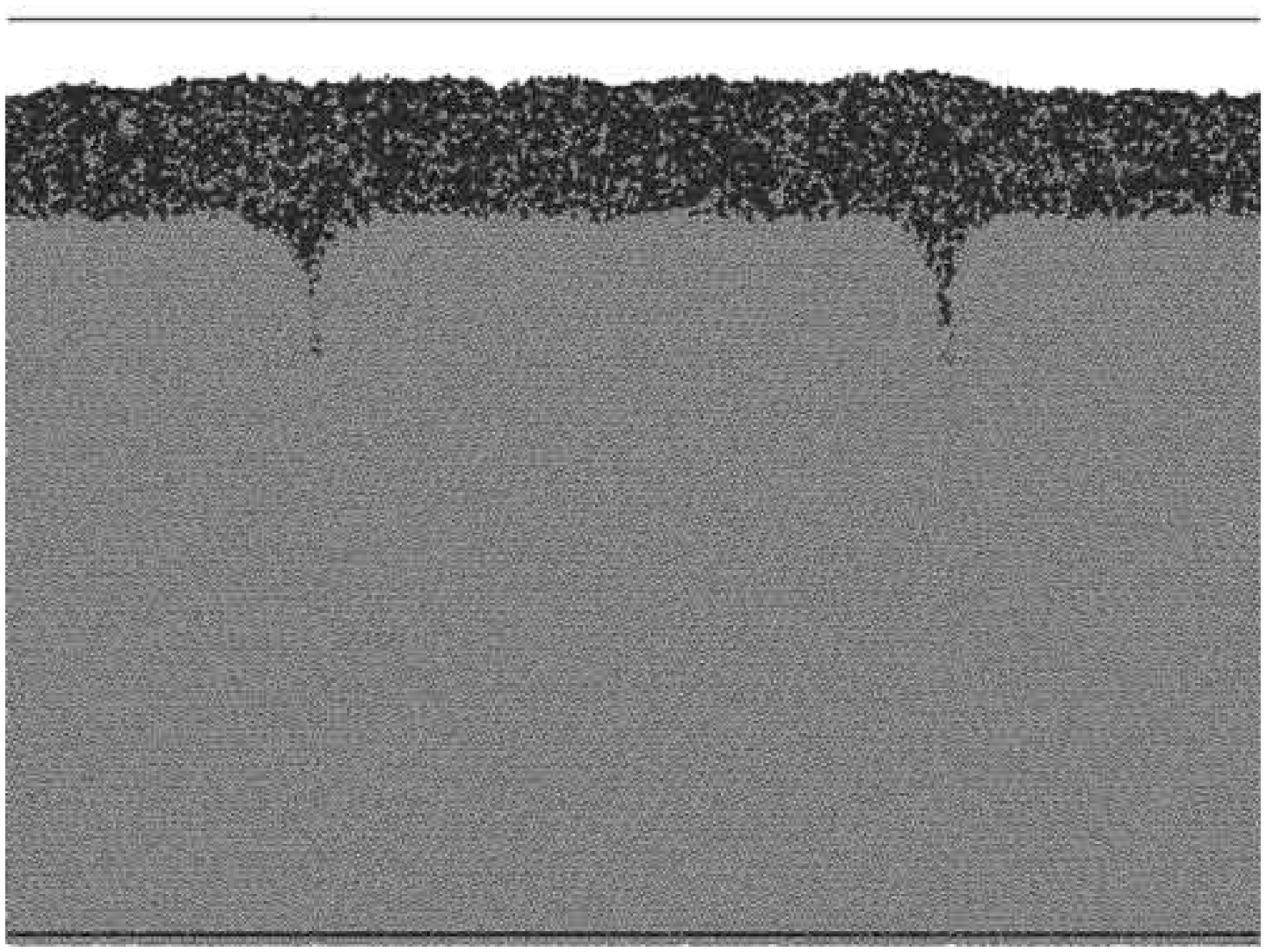}
\hspace*{0.1cm}
\includegraphics[width=0.32\textwidth]{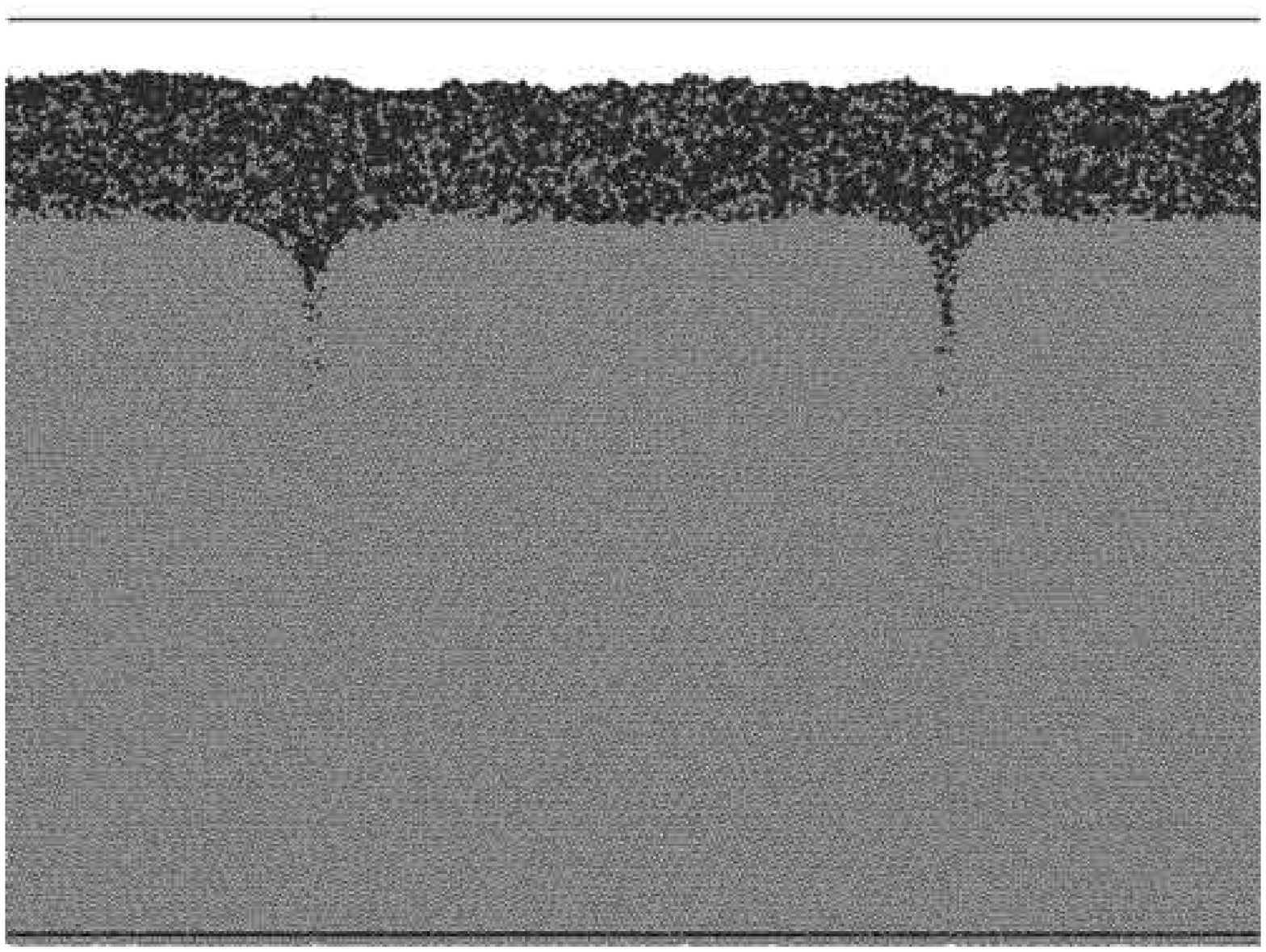}
\newline
\includegraphics[width=0.32\textwidth]{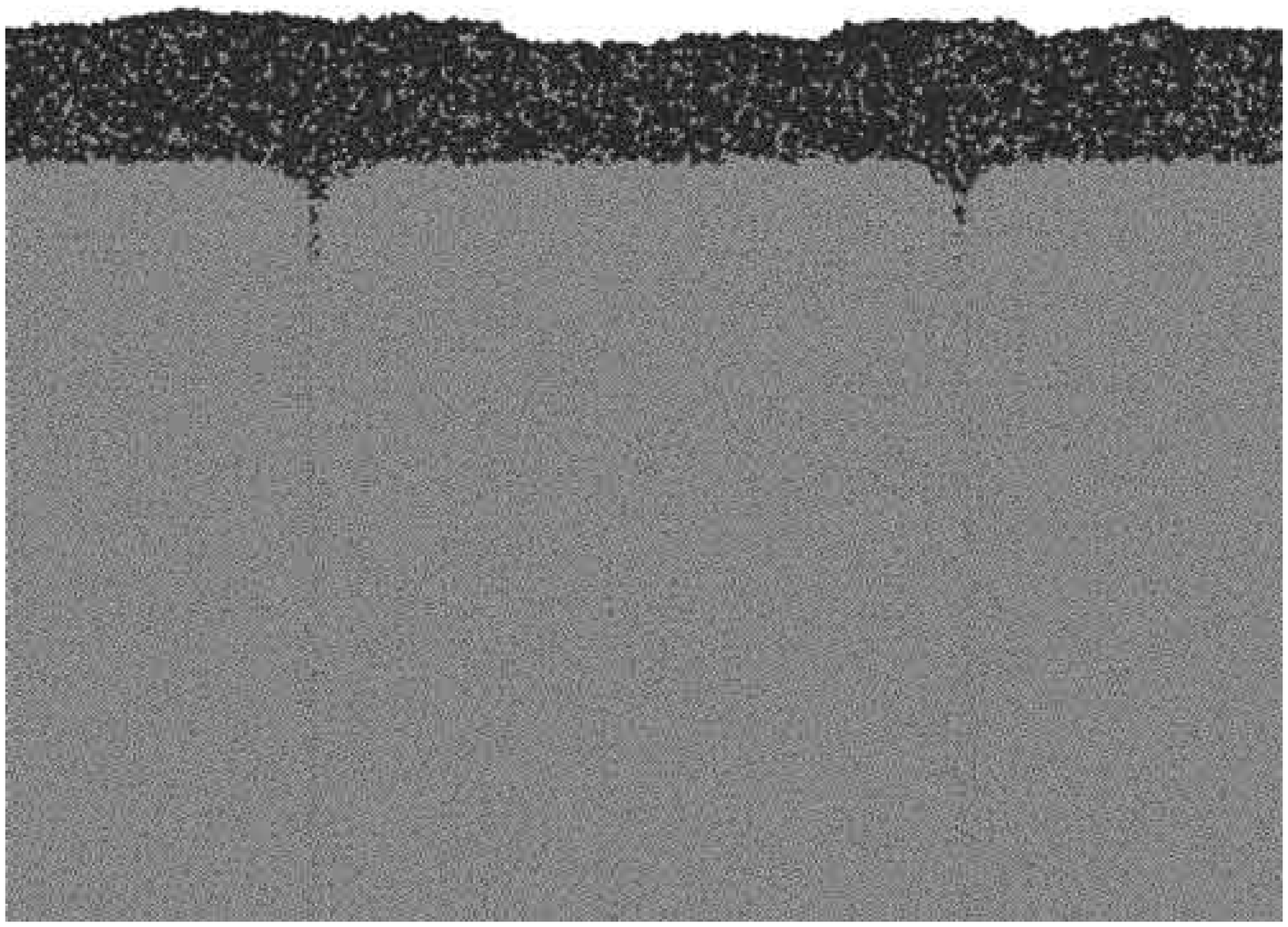}
\hspace*{0.1cm}
\includegraphics[width=0.32\textwidth]{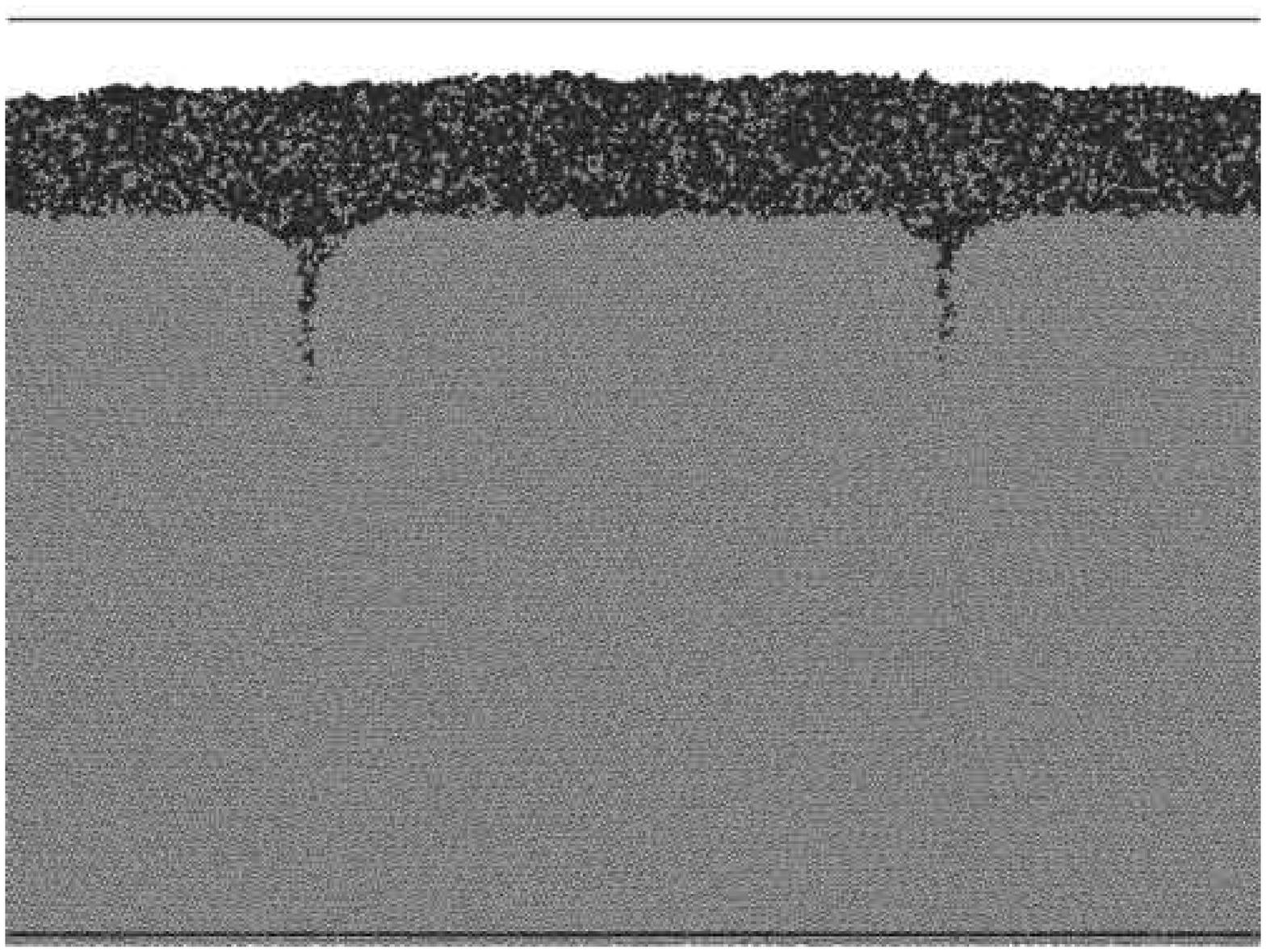}
\hspace*{0.1cm}
\includegraphics[width=0.32\textwidth]{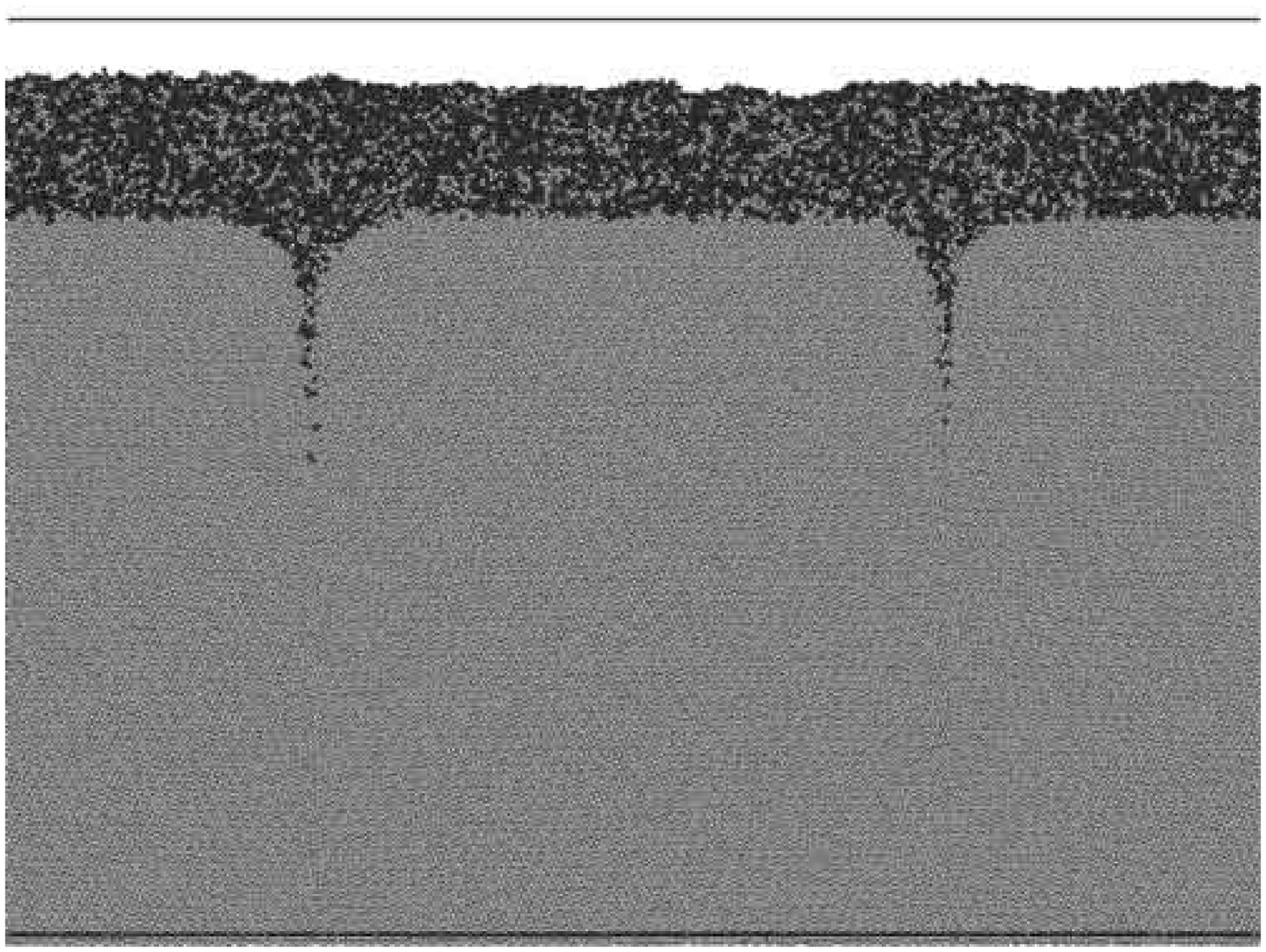}
\newline
\includegraphics[width=0.32\textwidth]{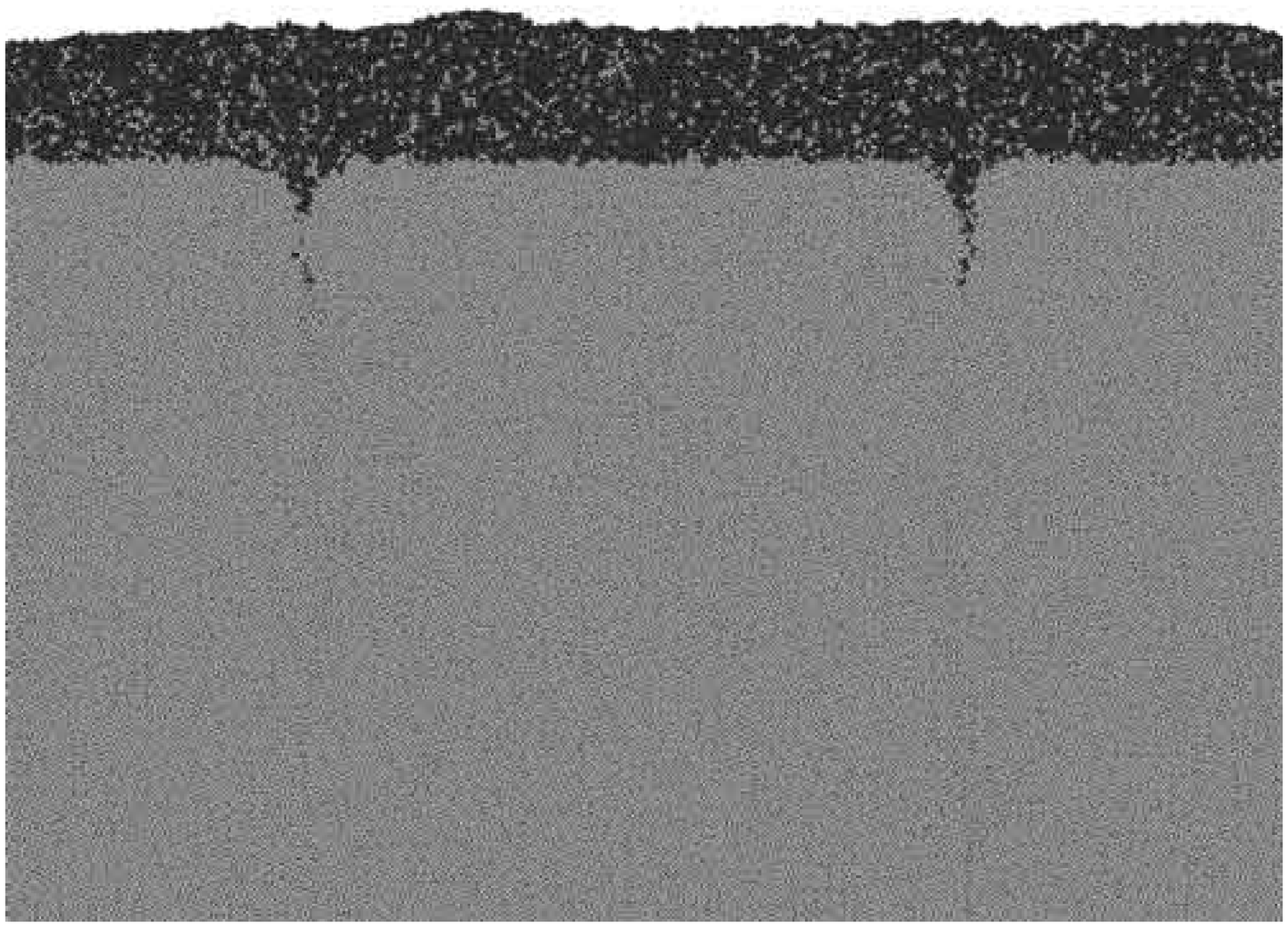}
\hspace*{0.1cm}
\includegraphics[width=0.32\textwidth]{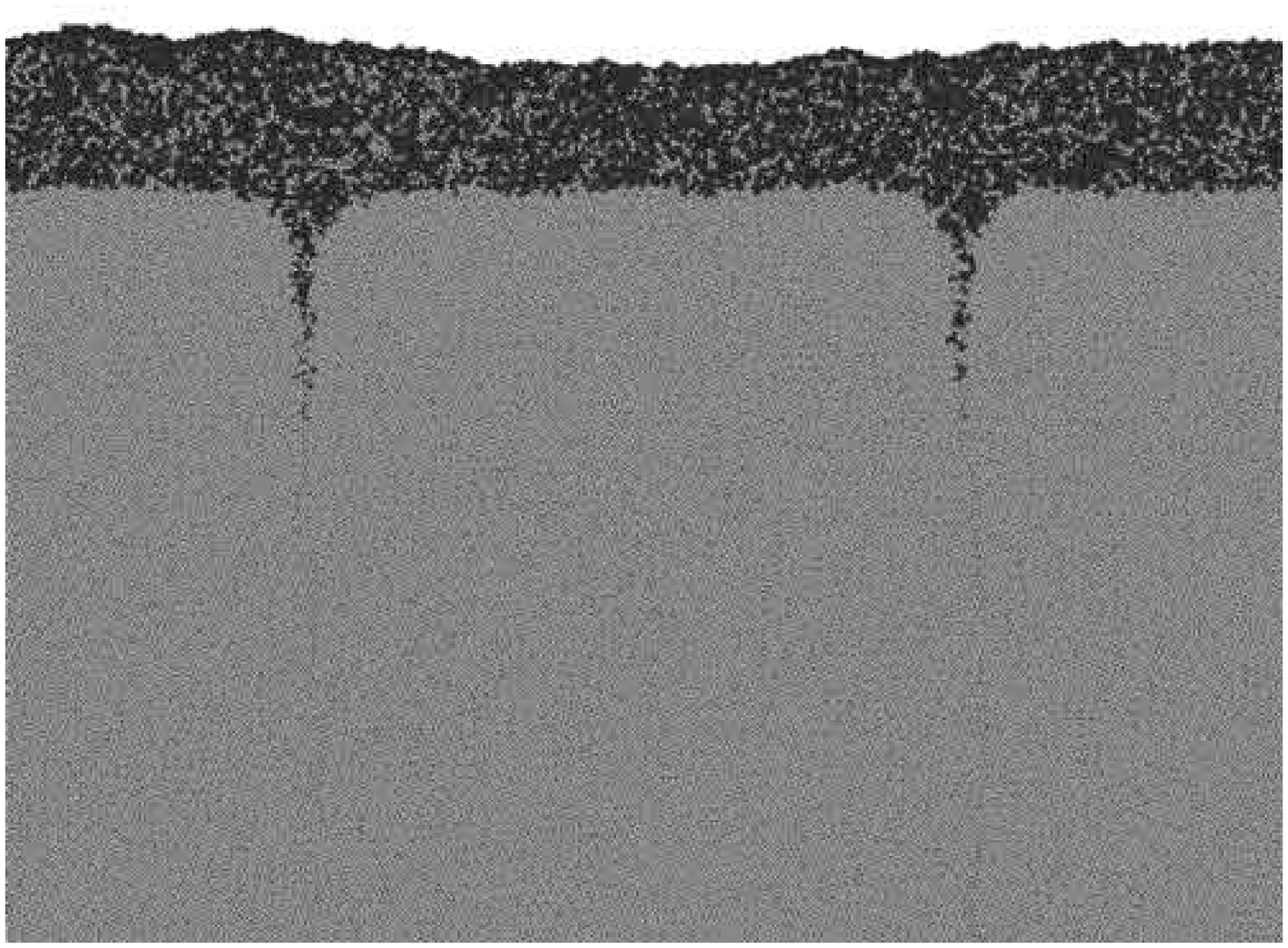}
\hspace*{0.1cm}
\includegraphics[width=0.32\textwidth]{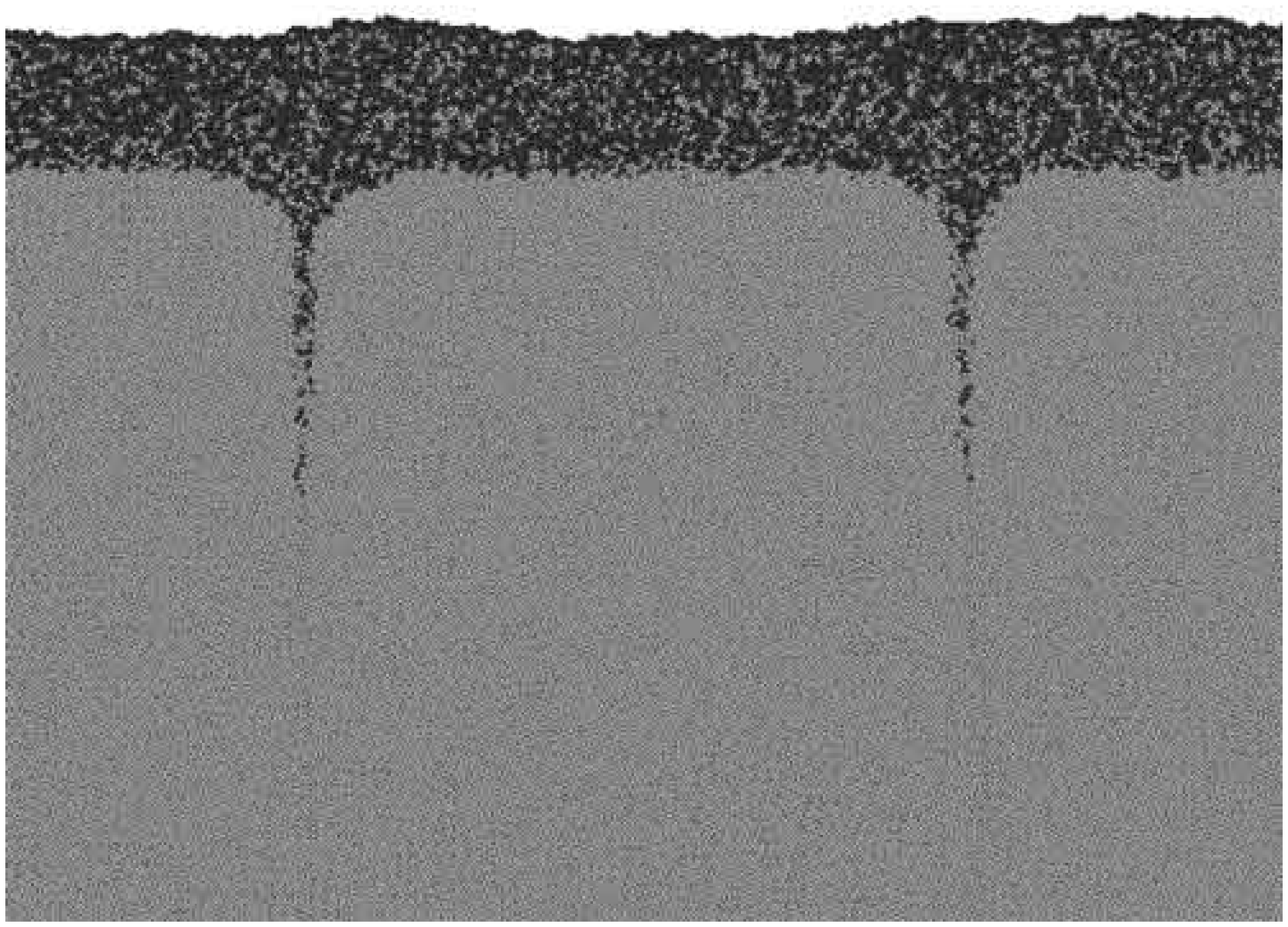}
\newline
\hspace*{0.1cm} (a) $\sigma_{xx}=0$ MPa \hspace*{3.2cm} (b) $\sigma_{xx}=250$ MPa \hspace*{3.1cm} (c) $\sigma_{xx}=500$ MPa \hspace*{0.0cm}
\caption
{ \label{fig:LME_sym}
Atomic scale images of liquid metal penetration in an Al bicrystal in contact with liquid Ga at $t=$ 10, 30, and 50 ns (from top to bottom) at constant strains corresponding to applied stresses $\sigma_{xx} $ of (a) 0, (b) 250, and (c) 500 MPa.  The atoms shown in grey represent Al atoms and those in black are Ga. 
}
\end{figure*}

To investigate the wetting of a grain boundary in Al in contact with liquid Ga under the influence of external stresses, we performed a series of MD simulations using the geometry of Fig.~\ref{fig:Geometry} at $T=600$ K.  Figure~\ref{fig:LME_sym} (a) shows the Ga penetration along the $\Sigma 5~ 36.9 \,^{\circ} /[010]$ symmetric tilt boundaries in the absence of an applied strain.  At the beginning of the simulation (within less then 10 ns), some Al atoms near the grain boundaries dissolve into the liquid Ga and liquid grooves form at the intersections of the grain boundaries and the solid-liquid interfaces.  Note that at $T = 600$ K, the solubility of Al in liquid Ga predicted by the phase diagram is $\sim 30$  mole percent.  The amount of Al that can dissolve into the liquid is limited by the finite quantity of liquid Ga present in the system.  As the liquid phase approaches saturation, the shape of the solid-liquid interface evolves very slowly and the rate of growth of the liquid groove decreases.  However, below the root of the liquid groove (i.e., where the liquid meets the grain boundary), Ga atoms continue to penetrate into the grain boundaries.  Examination of the simulation cell at the atomic level shows that the Ga distribution appears to be  diffusion-like rather than appearing as a liquid that abruptly terminates at a crack tip.  This is Ga penetration along the grain boundary.  Overall morphological and composition equilibrium is not yet obtained during the course of our 70 ns simulations.

Figures~\ref{fig:LME_sym}(b) and (c) show the formation of liquid grooves and Ga penetration at constant strains of about 0.65 $\%$ and 1.3 $\%$, respectively.  The imposed strains correspond to uniaxial stresses of $\sigma_{xx} \approx 250$ MPa and $\sigma_{xx} \approx 500$ MPa (i.e., normal to the nominal grain boundary plane), respectively, as  described in section~\ref{sec:sublevel_23LMEsimul}.  Although the liquid groove shape and wetting angle are nearly the same in Figs.~\ref{fig:LME_sym}(a)-(c), Ga penetration into the grain boundaries at the root of the liquid groove was greatly promoted by the application of the strain.

We also performed similar simulations for several other types of grain boundaries and found that the Ga penetration behavior and the effect of applied stresses can be very sensitive to grain boundary crystallography.  For example, we were unable to see any grain boundary wetting within our 50 ns simulations in low angle grain boundaries or the $\Sigma 5~ 36.9 \,^{\circ} /[100]$ symmetric twist grain boundary.  Since the $\Sigma 5~ 36.9 \,^{\circ} /[010]$ symmetric tilt boundary is a well-studied grain boundary in the literature and exhibits remarkable rate of Ga penetration in the simulations, we limit our focus to this grain boundary in the remainder of this paper.

\begin{figure}[!tbp]
\includegraphics[width=0.43\textwidth]{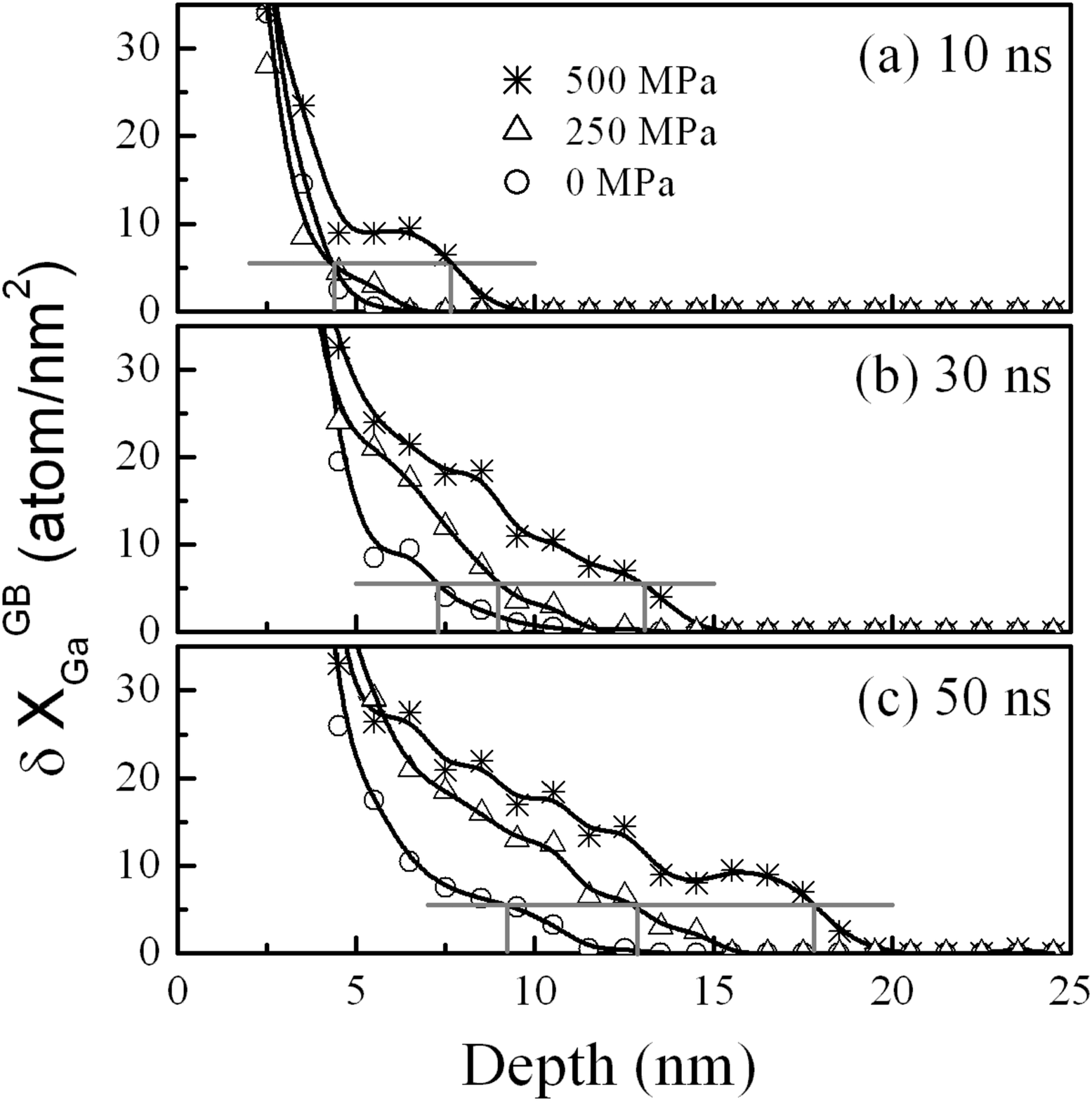}
\caption
{ \label{fig:GaProfile}
Ga penetration ($\delta X_{Ga}^{GB}$) profiles along the symmetric tilt grain boundaries with different applied stresses at (a) $t=10$ ns, (b) $t=30$ ns, and (c) $t=50$ ns.  The solid lines are cubic spline-fits and the vertical bars provide a measure of the Ga penetration depth, arbitrarily defined as the depth at which the Ga concentration is 6 atoms/nm$^2$ (solid horizontal line).
}
\end{figure}

Figure~\ref{fig:GaProfile} shows the effect of applied stress on Ga penetration rates.  In this figure, we plot the product of the grain boundary width $\delta$ and the Ga concentration $X_{Ga}^{GB}$ (Ga atoms per volume) versus distance along the symmetric tilt boundaries.  (The Ga concentration is measured in thin slices through the sample that are perpendicular to the nominal grain boundary.)  At the initial stage of liquid groove formation ($t<5$ nm), there is little effect of applied stress on the penetration profile.  This regime is dominated by dissolution.  We can arbitrarily define the Ga penetration depth by noting the depth at which the Ga concentration exceeds a fixed value at each time.  We mark the depth at which the grain boundary Ga concentration $\delta X_{Ga}^{GB}$ exceeds about half a monolayer ($\sim 6$ atoms/nm$^2$) at each time in Fig.~\ref{fig:GaProfile}.  As the grooves deepens, stress facilitates the rate of grain boundary penetration.  For example, Ga penetrates at least 10 nm along the grain boundaries between $t=$ 10 and 50 ns when an applied stress is present, but less than 5 nm when no stress is applied.  
%

\begin{figure}[!tbp]
\includegraphics[width=0.42\textwidth]{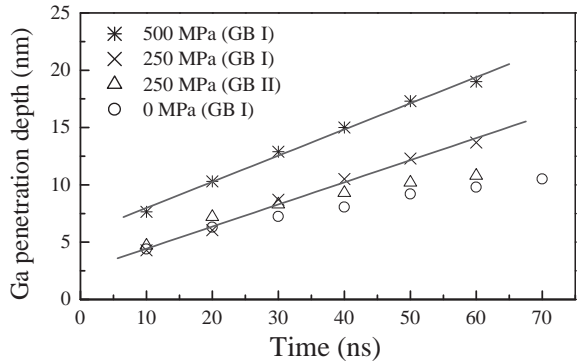}
\caption
{ \label{fig:PenetrationRate}
Ga penetration depth versus time with different applied stresses.  This depth is the distance from the surface where the Ga concentration $\delta X_{Ga}^{GB}=6$ atoms/nm$^2$ (see the horizontal line in Fig.~\ref{fig:GaProfile}). The two sets of data at 250 MPa correspond to two different grain boundaries in the same simulation.  
}
\end{figure}

We plot the Ga penetration depth $L$ versus time $t$ in Fig.~\ref{fig:PenetrationRate}.  In the absence of an applied stress, the rate at which Ga penetrates down the grain boundary (slope in Fig.~\ref{fig:PenetrationRate}) decreases with time.  However, when a stress is applied, the rate of Ga penetration appears to be nearly independent of time (the data in Fig.~\ref{fig:PenetrationRate} fall on straight lines).  [In the lower applied stress ($\sim 250$ MPa) case, only one grain boundary (GB I) shows a time-independent Ga penetration rate.  This is likely attributable to the fact that for the present very small grain size, Ga penetration along one grain boundary relieves the stress at the other grain boundary.]  Clearly, stress changes the fundamental nature of Ga penetration down grain boundaries in Al.  The time independence of the Ga penetration rate shows that the Ga is not simply undergoing random walk diffusion on the grain boundary ($L \propto t^{1/2}$) nor by normal grain boundary grooving~\cite{Vogel:GBInstability1971} ($L \propto t^{1/3}$ or $L \propto t^{1/4}$ for bulk or surface diffusion control, respectively), but is strongly driven ($L \propto t$).  The Ga penetration rate increases with the magnitude of the applied tensile stress and increasing grain size $d_{GB}$.

\section{\label{sec:level_4GBDiffSeg} Effects of Stress on Grain Boundary Properties}

\begin{figure}[!tbp]
\includegraphics[width=0.42\textwidth]{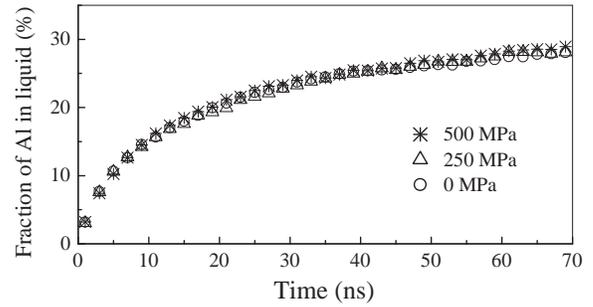}
\caption
{ \label{fig:Concentrationvstime_asy}
Fraction of Al in the liquid versus time for simulations performed under different applied stresses.
}
\end{figure}

To investigate the origin of the stress effect, we analyzed several, potentially relevant, physical properties associated with the grain boundary as a function of applied stress. Figure~\ref{fig:Concentrationvstime_asy} shows the time evolution of the mole fraction of Al in the liquid phase in the  MD simulations of Ga penetration reported above.  This quantity is related to the rate of Al dissolution (primarily) at the grain boundary.  The mole fraction of Al in the liquid increases with time from near zero, but appears to saturate at late times during the MD simulations.  The equilibrium solubility of Al in liquid Ga predicted by the phase diagram calculation should be $\sim 30$  mole percent.  This is very close to that observed by the end of the 70 ns simulations.  However, the three curves in Fig.~\ref{fig:Concentrationvstime_asy} are nearly indistinguishable from each other.  This demonstrates that stress has little effect on the rate or magnitude of Al dissolution.  This may not be surprising since the stress (normal traction) at the solid-liquid interface should be zero even in the presence of an applied stress.  Furthermore, application of a 500 MPa uniaxial stress should only lower the melting point by $\sim 1^\circ$ and increase the solubility of Ga in Al by much less than $1\%$ (based on the increase in the energy of the solid phase with strain and on the phase diagram).  Therefore, we cannot attribute the stress-effect to stress-driven changes in the solubility of Al in liquid Ga.  

\begin{figure}[!tbp]
\includegraphics[width=0.42\textwidth]{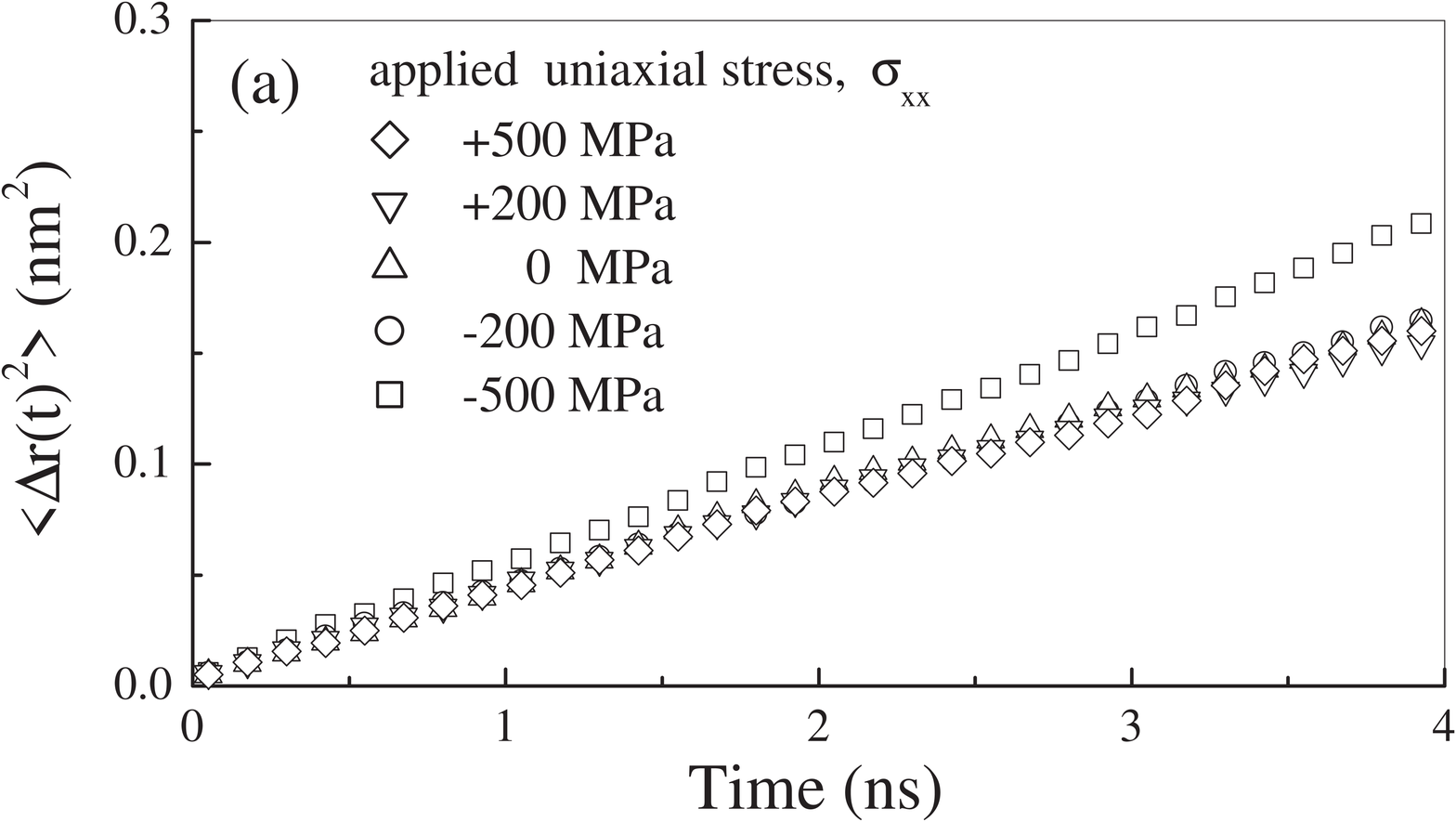}  \vspace{0.2cm}
\includegraphics[width=0.42\textwidth]{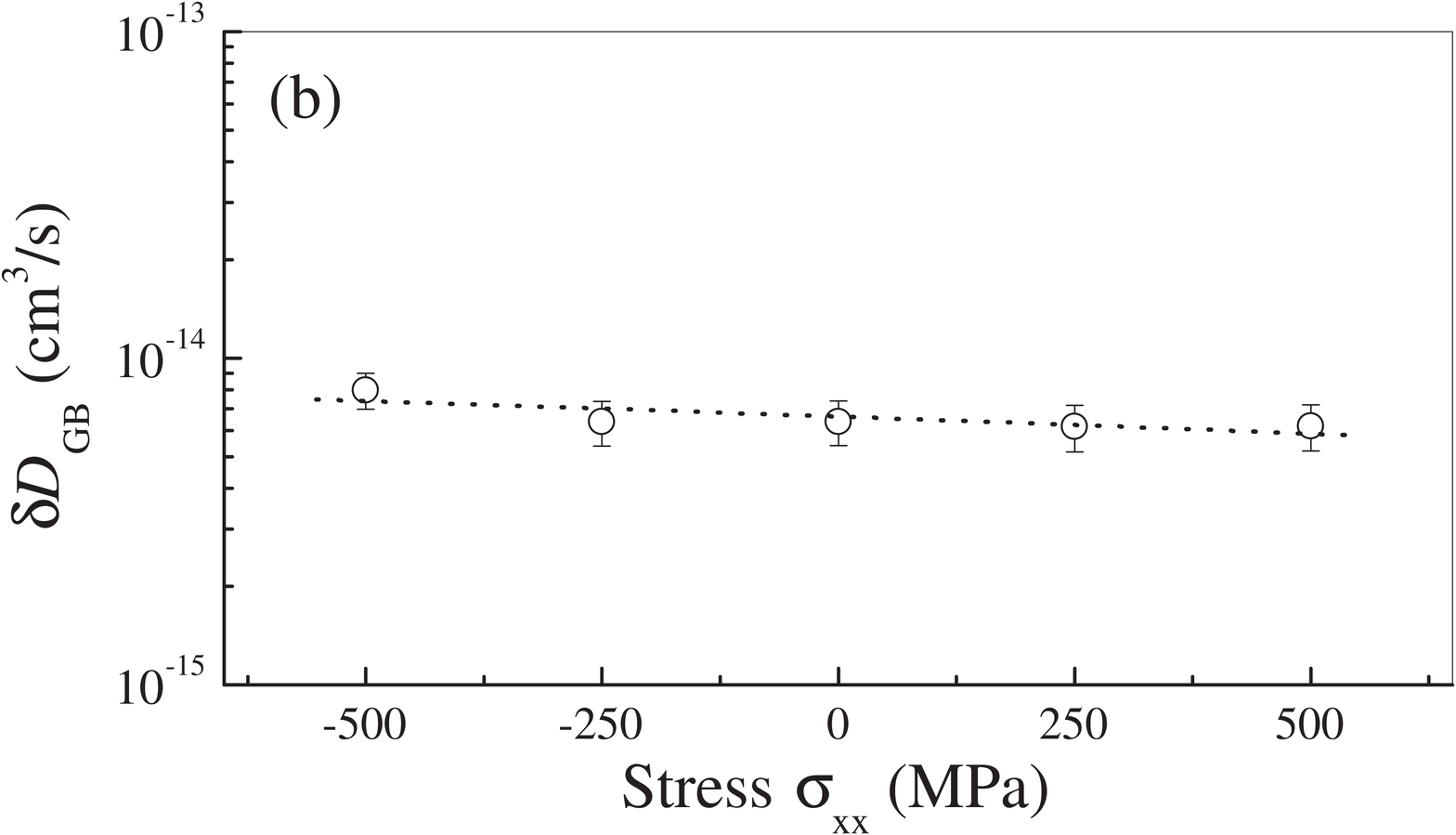}  
\caption
{ \label{fig:Dgb}
Grain boundary thickness - self-diffusivity product for the $\Sigma 5~ 36.9 \,^{\circ} /[010]$ symmetric tilt boundary at 600 K. (a) The mean-square displacement of the Al atoms in the grain boundary at different uniaxial stresses as a function of time.  (b) The logarithm of the grain boundary thickness - self-diffusion coefficient as a function of the applied stress.
}
\end{figure}

Another possible explanation for the role of an applied stress on the rate of Ga penetration down grain boundaries in Al may be associated with the effect of stress on Ga diffusion along the grain boundaries in Al.  Unfortunately, it is difficult to measure the diffusivity of Ga along an Al grain boundary because the number of Ga tracer atoms would be too small to achieve sufficient statistical accuracy at low Ga concentration and, at high concentrations, the grain boundary pulls apart in the presence of an applied stress (at $600$ K).  If the impurity diffusivity is correlated with the self-diffusivity via a vacancy mechanism,~\cite{Balluffi:GBdiffusion} the Al self-diffusivity along the grain boundary is relevant.  Figure~\ref{fig:Dgb}(a) shows the in-plane mean-squared displacement $\langle\Delta r(t)^2\rangle(=\langle\Delta x(t)^2\rangle+\langle\Delta y(t)^2\rangle)$ for Al atoms in the $\Sigma 5$ symmetric tilt boundary as a function of time at different uniaxial stresses, $\sigma_{xx}$, in an effectively infinite Al bicrystal (i.e., no Ga/no liquid).  The mean squared displacement is a linear function of time.  Figure~\ref{fig:Dgb}(b) shows the the boundary width ($\delta D_{GB}$) - self-diffusion coefficient product obtained from the slopes of the curves in Fig.~\ref{fig:Dgb}(a).  These results suggests that the grain boundary self-diffusivity is very nearly independent of applied stress (to within the accuracy of the simulations).  Therefore, the effect of stress on grain boundary diffusion is an unlikely source for the difference in Ga penetration rates observed with different applied stresses.

\begin{figure}[!tbp]
\includegraphics[width=0.40\textwidth]{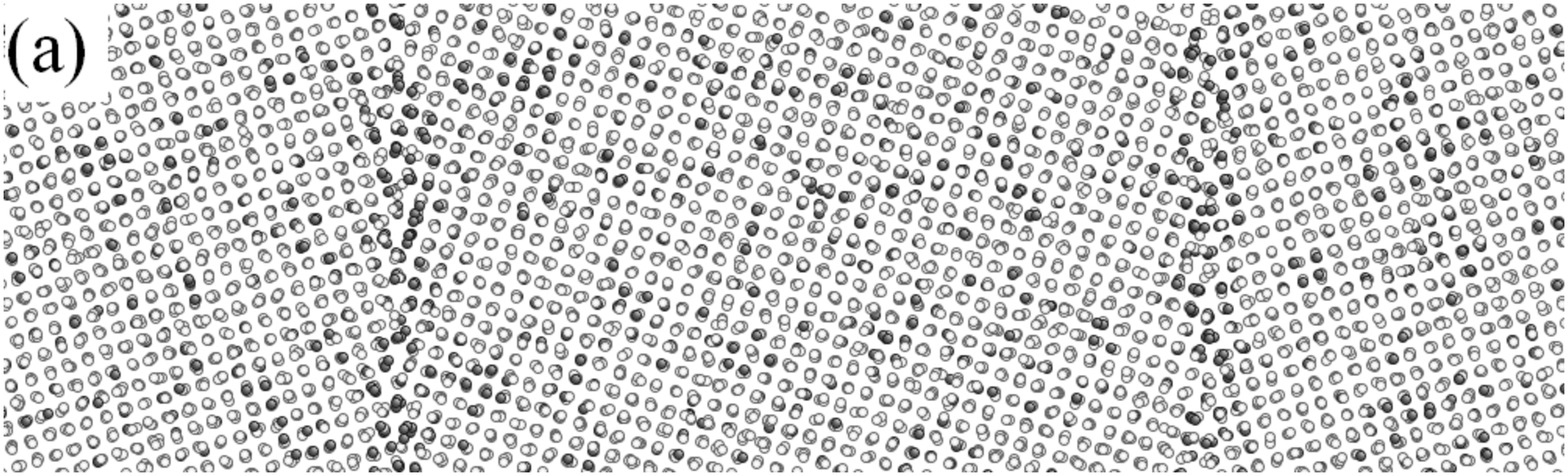}
\includegraphics[width=0.40\textwidth]{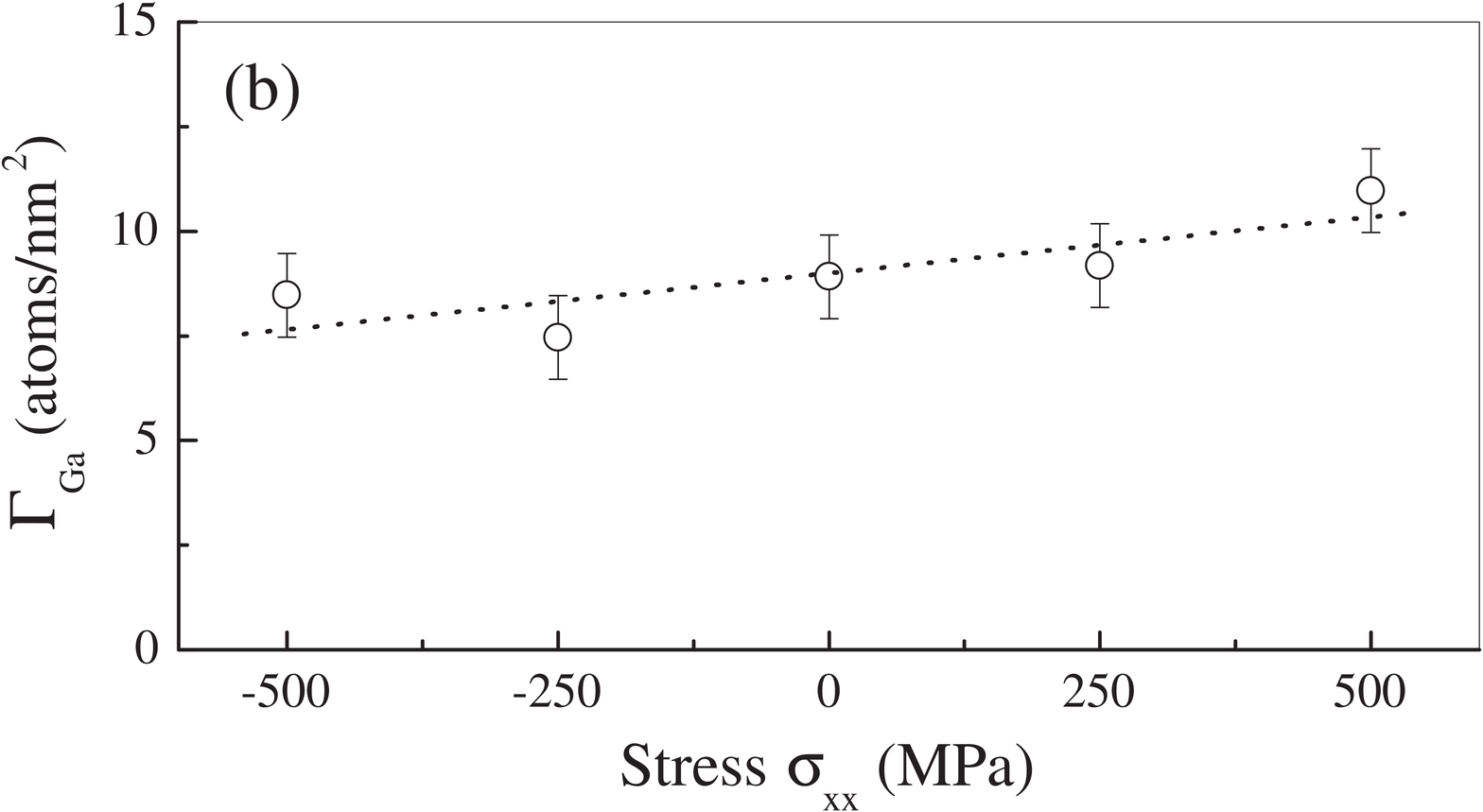}
\caption
{ \label{fig:GBseg}
(a) An atomic scale image from a semi-grand canonical Monte-Carlo simulation of a $\Sigma 5~ 36.9 \,^{\circ} /[010]$ symmetric tilt boundary in the absence of an applied stress at 600 K.  The white and black circles represent Al and Ga atoms, respectively.  (b) The Gibbsian grain boundary excess $\Gamma_{Ga}$ of Ga versus the applied stress.
}
\end{figure}

Next, we investigated the effect of stress on the segregation of Ga to the grain boundary.  To this end, we perform a series of Monte Carlo simulations of a bicrystal (no liquid) in a semi-grand canonical ensemble where we fix the chemical potentials of Al and Ga such that the bulk concentration of Ga is $\sim 10\%$ at $T=600$ K.  To simulate segregation in the presence of a uniaxial stress, we adopted the isotension-isothermal ensemble.~\cite{Wojciechowski:IsotensionMC}  Figure~\ref{fig:GBseg}(a) shows the atomic structure of an Al-Ga bicrystal with symmetric tilt grain boundaries in the absence of an applied stress.  Clearly, Ga segregation to the grain boundary does occur and leads to a relatively disordered grain boundary structure at $T=600$ K.  We measure the Gibbsian grain boundary excess $\Gamma_{Ga}$ of Ga at the grain boundary from the concentration profile for several applied stress, as shown in Fig.~\ref{fig:GBseg}(b).  The grain boundary excess is the quantity of solute present per unit area of flat interface in excess of the quantity of solute that would be present if there were no grain boundary (i.e., the bulk concentration).  The grain boundary may be defined as $\Gamma_{Ga} = (N_{total}-C_{bulk}V_{total})/A_{GB}$, where $N_{total}$ is the total number of solute atoms in the cell volume $V_{total}$, $C_{bulk}$ is the bulk concentration of solute, and $A_{GB}$ is the grain boundary area in the volume.  As shown in Fig.~\ref{fig:GBseg}(b),  applied stresses have  little impact on the degree to which Ga segregates to grain boundaries in Al.

\section{\label{sec:level_5StressProfile} Distribution of Stresses}

Since stress has little effect on many of the basic physical parameters that could, in principle, be responsible for the stress effect in LME, we examine the distribution of the stress within our Al-Ga bicrystal system.  
\begin{figure*}[!tbp]
\includegraphics[width=0.98\textwidth]{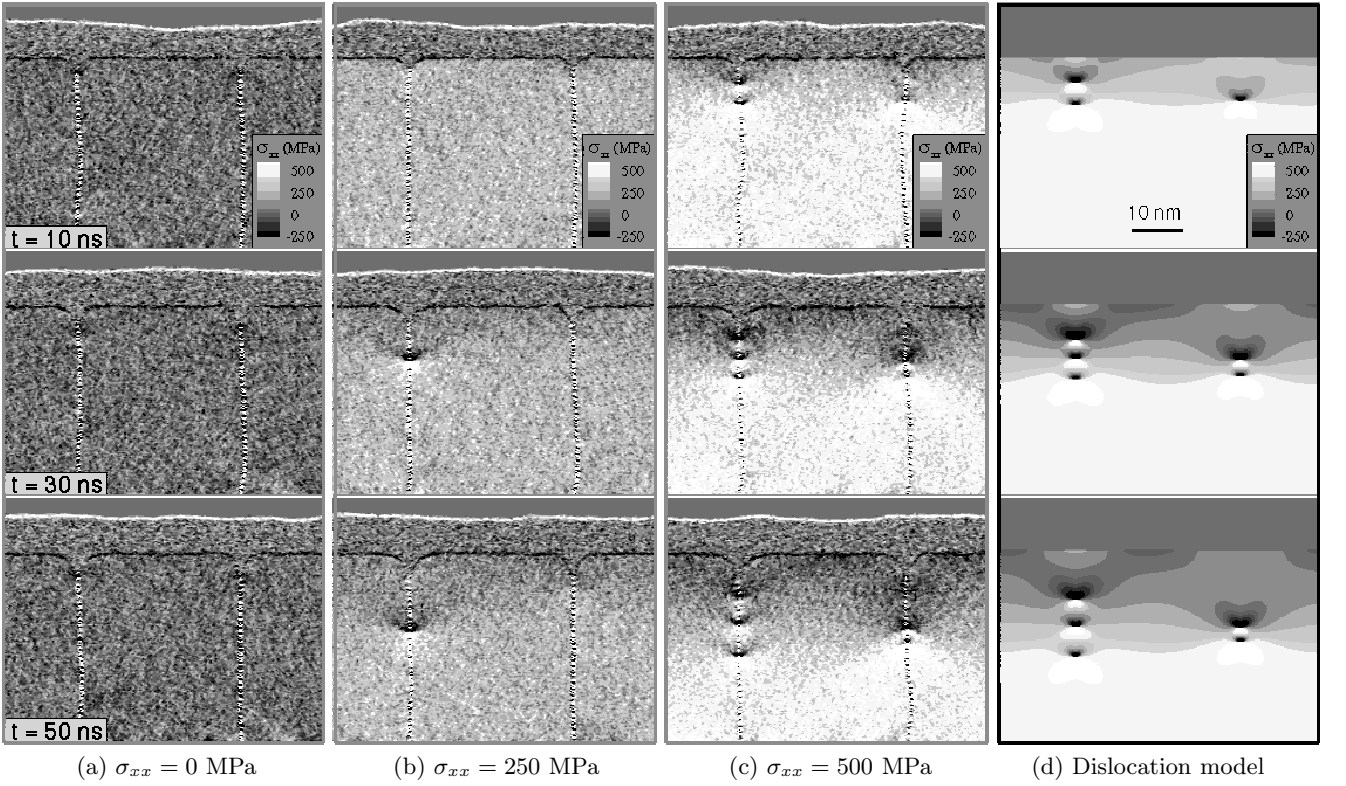}
\newline
\hspace*{0.0cm} (a) $\sigma_{xx}=0$ MPa \hspace*{1.6cm} (b) $\sigma_{xx}=250$ MPa \hspace*{1.5cm} (c) $\sigma_{xx}=500$ MPa \hspace*{1.1cm} (d) Dislocation model
\caption
{ \label{fig:StressField}
Stress distributions ($\sigma_{xx}$) at $t=$ 10, 30, and 50 ns (from top to bottom) for simulations performed at $T=600$ K at constant strains corresponding to applied stresses $\sigma_{xx} $ of (a) 0, (b) 250, and (c) 500 MPa.  (d) The stress distribution predicted from the dislocation model (see the text) is shown for comparison with (c).  
}
\end{figure*}
Figure~\ref{fig:StressField} shows the time evolution of the stress distribution for $\sigma_{xx}$ within the system at constant strains of 0, $0.65 \%$ ($\sigma_{xx} \approx 250$ MPa), and $1.3 \%$ ($\sigma_{xx} \approx 500$ MPa).  Figure~\ref{fig:StressField}(a) shows that in the absence of an applied strain, the stresses in the system are small and random.  However, when a strain is applied, we note the formation of one [Fig.~\ref{fig:StressField}(b)] or more [Fig.~\ref{fig:StressField}(c)] stress concentration patterns at the grain boundary.  These patterns consist of a dark (large compressive) region above a light (large tensile) region.  This stress patterns are suggestive of stress patterns expected for edge dislocations with a Burgers vector perpendicular to the boundary plane.  

We can use linear elastic theory to predict the stress field associated with such an edge dislocation in a linear elastic half-space using the approach of Head~\cite{Head:EdgeDislStress}.  This stress field $\sigma_{xx}$ corresponding to a set of edge dislocations with Burgers vector $b$ parallel to the free surface $z$=0, located at points $(\xi_i, \eta_i)$ in the $xz$-plane (see Fig.~\ref{fig:Geometry}) is given by  
\begin{widetext}
\begin{eqnarray}
\label{eq:EdgeDislocationStress}
\sigma_{xx} = \frac {E b} {4 \pi (1-\nu^2)} 
\sum_{i} \biggl[
  \frac {(z-\eta_i) \{ (z-\eta_i)^2 + 3(x-\xi_i)^2 \} } { \{ (z-\eta_i)^2 + (x-\xi_i)^2 \}^2 } 
- \frac {(z+\eta_i) \{ (z+\eta_i)^2 + 3(x-\xi_i)^2 \} } { \{ (z+\eta_i)^2 + (x-\xi_i)^2 \}^2 } 
\nonumber\\
~~~- 2\eta_i \frac {(z-\eta_i)(z+\eta_i)^3 - 6z(z+\eta_i)(x-\xi_i)^2 + (x-\xi_i)^4} { \{ (z+\eta_i)^2 + (x-\xi_i)^2 \}^3 } 
\biggr] ,  
\end{eqnarray}
\end{widetext}
where $E$ and $\nu$ are the Young's modulus and Poisson's ratio, respectively.  Figure~\ref{fig:StressField}(d) shows the predicted stress field corresponding to Fig.~\ref{fig:StressField}(c), where we have assumed that the positions of the dislocations in Fig.~\ref{fig:StressField}(d) are the same as those observed in Fig.~\ref{fig:StressField}(c).  In making this comparison, we have determined the magnitude of the Burgers vector in Fig.~\ref{fig:StressField}(d) by adjusting it to obtain best correspondence with the stress distributions in Fig.~\ref{fig:StressField}(c).  We find that the best fit is obtained for $|b| \approx 0.5$ \AA, which is much smaller than the Burgers vector of a lattice dislocation, $\sim 2.86$ \AA.  

No dislocations were formed in the case of zero applied stress [Fig.~\ref{fig:StressField}(a)] or at one of the two grain boundaries in the $\sigma_{xx}=250$ MPa [Fig.~\ref{fig:StressField}(b)].  It is interesting to note that it was only in exactly these two cases that the Ga penetration depth versus time plot (Fig.~\ref{fig:PenetrationRate}) were sub-linear (i.e., the penetration rate decreases with time).

Examination of Fig.~\ref{fig:StressField}(b) shows that a dislocation, once formed, ``climbs'' down along the grain boundary at a nearly constant rate.  Examination of Fig.~\ref{fig:StressField}(c) (larger strain) shows that the dislocations ``climb'' down the grain boundaries at the same constant rate as the single dislocation in Fig.~\ref{fig:StressField}(b) (low strain).  However, in this case, once the first dislocation has moved some distance from its point of origin (below the liquid groove root), second and third dislocations are nucleated one after another (boundary on the left in Fig.~\ref{fig:StressField}(c)), and ``climb'' down the grain boundaries too, leading to equally spaced dislocations that all move at the same rate.  Therefore, we can conclude that these special grain boundary dislocations only form above a critical applied strain/stress and ``climb'' down the grain boundaries at a constant rate that is independent of the magnitude of this strain.  Increasing applied strain simply results in the formation of more dislocations with shorter incubation time.  
\begin{figure*}[!tbp]
\includegraphics[width=0.98\textwidth]{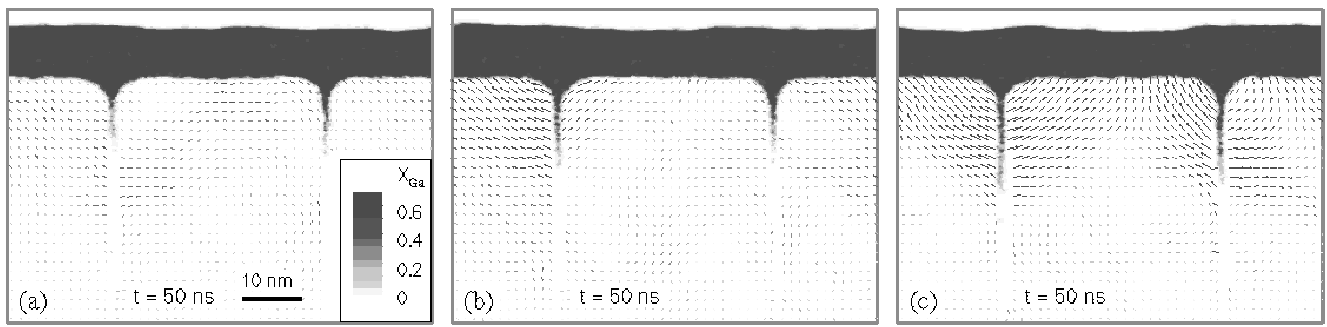}
\caption
{ \label{fig:Displacement}
Ga concentration profiles (contour plots of mole fraction $X_{Ga}$) and displacement fields, measured between t= 10 and 50ns, for the bicrystal with an applied stress of (a) 0, (b) 250, and (c) 500 MPa.  The displacement vectors correspond to averages over several hundred atoms and are shown magnified by a factor of ten for better resolution.  In order to limit consideration to displacements associated with elastic deformation of the solid, we excluded atoms for which the displacements exceeded 5 {\AA} (i.e., primarily diffusive hops) and, as a result, no arrows are plotted in the grain boundary region.
}
\end{figure*}

In order to understand the relaxation of the stresses, we also analyzed the atomic-level displacement fields.  Figure~\ref{fig:Displacement} shows the Ga concentration profiles at $t=50$ ns (contour plots) and the displacements in the solid that occurred between $t=$ 10 and 50 ns (i.e., the displacement vectors are measured as the atom positions at 10 ns to the positions of the same atoms at 50 ns) at applied strains of 0, $0.65 \%$ ($\sigma_{xx} \approx 250$ MPa), and $1.3 \%$ ($\sigma_{xx} \approx 500$ MPa).  

In the absence of an applied stress, only small displacements were observed and these displacements were distributed randomly over the entire solid (consistent with the stress distributions shown in Fig.~\ref{fig:StressField}(a)).  When a stress is applied, the atomic displacements are considerably larger than in the absence of an applied stress and a distinctive displacement vector pattern emerges [see Figs.~\ref{fig:Displacement}(b) and (c)].  The atomic displacements near the grain boundary groove tip are directed away from the grain boundary.  This is likely associated with  Ga atoms being inserted into the grain boundary to reduce the tensile stresses there.  The largest atomic displacements observed within the grains adjacent to where Ga has thoroughly penetrated the nearby grain boundary.  

Additional simulations, not shown here, demonstrate that doubling the grain size doubles the magnitude of the displacements.  This can be traced to the fact that the present simulations were performed under fixed-grip loading (to a particular bulk stress value).  This implies that doubling the grain size doubles the strain energy stored in the sample.  To relieve the same stress in a system with twice the grain size requires twice the grain boundary opening displacement.  This is consistent with the observation that increasing the grain size leads to a dramatic increase in the quantity of Ga at the grain boundary.~\cite{HoseokNam:LMEActa} 

\begin{figure}[!tbp]
\includegraphics[width=0.43\textwidth]{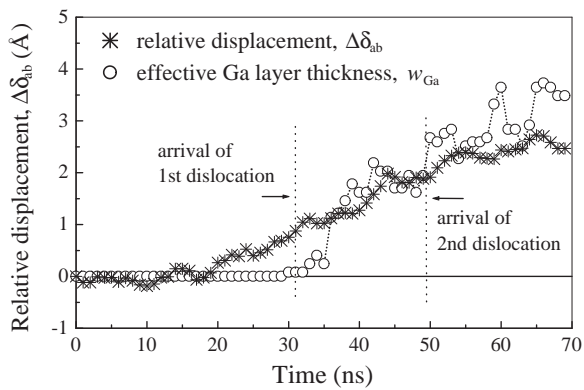}
\caption
{ \label{fig:GBseparation}
Relative displacement between two points in adjacent grains versus time with an applied stress of 500 MPa.  The distance between a pair of points was measured at $z=15$ nm below the initial surface (see Fig.~\ref{fig:Geometry}).  The effective Ga-rich layer thickness $w_{Ga}$ versus time measured at the same position is also shown.  
}
\end{figure}

We also analyzed the relative displacement between a pair of points in adjacent grains as an indicator of grain boundary opening; the relative displacement $\Delta \delta_{ab}(t) (=\delta_{ab}(t) - \delta_{ab}(0))$ is the change of the distance $\delta_{ab}(t)$ between the two points indicated in Fig.~\ref{fig:Geometry} at time $t$ with respect to the distance at zero time.  This displacement is a measure of the grain boundary opening distance.  This type of analysis was used earlier in experimental observations of LME (e.g., see the Al-Ga experiments in Ref.~\onlinecite{Pereiro-Lopez:AlGaPRL2005}).  Figure~\ref{fig:GBseparation} shows the relative displacement $\Delta \delta_{ab}(t)$ of a pair of points from adjacent grains and the effective Ga layer thickness $w_{Ga}$ as a function of time.  (The effective thickness of the Ga layer quantifies the quantity of Ga at the grain boundary and is defined as $w_{Ga}=N_{Ga}\Omega/A_{GB}$, where $N_{Ga}$ is the number of Ga atoms in a slab perpendicular to the grain boundary  times  the atomic volume of Ga  $\Omega$ and divided by the cross-sectional area of the slab $A_{GB}$.)  $\Delta \delta_{ab}$ and $w_{Ga}$ are shown as a function of time in Fig.~\ref{fig:GBseparation}, as measured at $z=15$ nm below the initial surface.  Clearly, the relative displacement $\delta_{ab}$ is initially zero then, at a finite time, gradually increases as Ga penetrates into the grain boundary.  The effective Ga layer thickness $w_{Ga}$ is also initially zero, but then increases abruptly at a time later than where the relative displacement begins to grow.  The abrupt increase in the effective Ga layer thickness occurs when a dislocation passes the measurement depth $z=15$ nm, as observed in Fig.~\ref{fig:StressField}(c).  Subsequent abrupt rises in $w_{Ga}$ correspond to the passage of additional dislocations. Interestingly, the effective Ga layer thickness is comparable to the grain boundary opening distance.  A similar evolution of $\Delta \delta_{ab}(t)$ and $w_{Ga}(t)$ occurs at points farther from the surface (not shown) at later times (set by the dislocation climb velocity).  Although the time and length scales are different, the shape of these curves from the simulations in Fig.~\ref{fig:GBseparation} are very similar to those measured experimentally by Ref.~\onlinecite{Pereiro-Lopez:AlGaPRL2005}.  
The following picture emerges.  Ga diffuses down the grain boundary leading to some grain boundary opening.  At a critical opening, a dislocation forms and starts to climb downward, trailing a Ga concentration tail behind it.  This Ga concentration at the boundary and the dislocation itself lead to some grain boundary opening ahead of the dislocation (and Ga concentration tail).   The grain boundary opening, dislocation and Ga tail move down the grain boundary at the same rate.  In the next section, we propose a mechanism by which all of this (i.e., LME) occurs.

\section{\label{sec:level_6Discuss} Discussion}

We performed a series of MD simulations of an Al bicrystal in contact with liquid Ga and examined the penetration of Ga along the Al grain boundaries in the presence and absence of an applied stress.  Even though the present simulations were relatively large, both in terms of the number of atoms modelled and simulation time, our atomistic simulation approach to LME, nonetheless, has some limitations.  For example, the simulations were only able to describe the first $\sim 70$ ns of the LME process and the extent of the liquid metal penetration was limited to $\sim 40$ nm.  This should be compared with many experiments in which thin (nanometer-thick) liquid films penetrate up to several hundred micrometers over time scales ranging from seconds to hours.  Because of these limitations, the stresses applied in the present simulation had to be much larger than those applied in experiment in order to observe LME.  The relatively small size of our simulation cells also precluded observations of significant plastic deformation (dislocation nucleation in small volumes is difficult and there were no pre-existing dislocations), even though plastic deformation would occur in macroscopic samples at this stress level.  However, the fact that significant plastic deformation does not occur during our simulations was fortuitous since experiments show that LME occurs in Al-Ga at stresses which are too small to cause macroscopic deformation of the Al. 

The molecular dynamics simulations can complement continuum models and experiments.  First, simulations can provide an atomic scale view of the dynamics of the system, grain boundary structure, and stresses and composition near the advancing liquid film layer front.  Such fundamental, high resolution data are rarely accessible in experiments: for example, although the synchrotron radiation X-ray microradiography~\cite{Ludwig:MSEA2000, Pereiro-Lopez:AlGaBiPoly2003, Pereiro-Lopez:AlGaPolycrystal2004, Ludwig:AlGaBicrystal2005, Pereiro-Lopez:AlGaPRL2005, Pereiro-Lopez:AlGaInvasion2006, Pereiro-Lopez:AlGaStress2006} is very useful for {\em in-situ} observation of sub-micrometer Ga wetting layers, the very early stages of the wetting process involves a Ga layers that are only a few monolayers thick and are not visible with this technique (typically, a layer thickness of at least $\sim 5$ nm is required for detection).  The molecular dynamics simulations provide a unique tool for understanding which materials properties are important in LME and for identifying which processes are rate-controlling.  This is accomplished by careful design of idealized set of rigidly maintained conditions to emphasize certain effects and exclude competing factors which may influence the LME.  While no interatomic potential is perfect, the EAM potentials for the Al-Ga alloys were shown to successfully reproduce many of the important solid-liquid properties in this alloy system.  To complement the known behavior of these potentials, we also performed an analysis of several other properties of this system that may affect LME, including grain boundary diffusivity and the tendency for Ga grain boundary segregation.  This gave us the ability to exclude certain models for LME from further consideration.

Our simulation results may be directly compared with many experimental observations in the literature.  This is, in part, associated with the fact that Al-Ga is one of the most widely experimentally studied LME systems (both polycrystals and bicrystals).  Among these, the recent series of TEM studies by \citeauthor{Hugo:AlGaTEM1998},~\cite{Hugo:AlGaTEM1998, Hugo:AlGaBicrystalTEM1999, Hugo:AlGaAtomicModelTEM2000} and synchrotron radiation micro-radiography studies by Ludwig and Pereiro-Lopez {\em et al.}~\cite{Ludwig:MSEA2000, Pereiro-Lopez:AlGaBiPoly2003, Pereiro-Lopez:AlGaPolycrystal2004, Ludwig:AlGaBicrystal2005, Pereiro-Lopez:AlGaPRL2005, Pereiro-Lopez:AlGaInvasion2006, Pereiro-Lopez:AlGaStress2006} are of particular interest because they are both quantitative and provide microscopic observations.  Although these experimental techniques are very different, both studies report a consistent set of features.

Although the time and length scales of the simulations and experiments differ, we demonstrated that our simulations were able to capture many of the experimentally observed trends in liquid Ga penetration of grain boundaries in Al.  Experiments showed that the application of tensile stresses (as small as a few MPa) to an Al polycrystal sample, drastically increased the liquid metal penetration rate.~\cite{Nicholas:LMEreview} Recent synchrotron radiation X-ray micro-radiographic experiments~\cite{Pereiro-Lopez:AlGaStress2006} showed that the penetration behavior of liquid Ga along two different types of symmetrical tilt  bicrystals of Al was greatly facilitated by the application of a tensile stress of only $\sim 5$ MPa.
Our simulations also showed that an applied stress significantly increases the rate of liquid metal penetration, although the applied stress level used in the simulations was approximately 100 times larger than that used in the experimental studies ($\sim 500$ versus $\sim 5$ MPa).  As in the macroscopic experiments,~\cite{Popovich:LMEreview, Nicholas:LMEreview} the simulations demonstrated that the penetration rate and thickness of the Ga-rich layer increases with increasing grain size~\cite{HoseokNam:LMEActa} (under fixed grip conditions).  Since grain sizes in most real materials are much larger than those used in the simulations ($\sim 1$ mm versus $\sim 33$ nm, a factor of $\sim 3 \times 10^4$), the difference in the magnitude of the applied stress necessary to facilitate the penetration rate in the simulations is understandable.  

It is clear that, even in the absence of an applied stress, many experimental samples may be subject to residual stresses resulting from sample fabrication, processing, polishing, or gripping.  Ludwig and Pereiro-Lopez {\em et al.}~\cite{Ludwig:AlGaBicrystal2005, Pereiro-Lopez:AlGaPolycrystal2004} demonstrated that residual stress, introduced by sample preparation, can affect liquid metal penetration rates in both Al bicrystals and polycrystals.   In their experiment, susceptibility to grain boundary penetration was strongly influenced by the method in which the sample was gripped.  Two different gripping procedures were used in their experiments: ``gently screwing'' together between metal plates and ``gluing'' the Al sample to Cu supports using silver paint.  The screwed samples exhibited larger liquid film propagation and thickening rates, as well as a larger probability that specific grain boundaries (in Al bicrystals) could be penetrated at all.  Although the magnitude of these effects was not accurately quantified, the sensitivity of grain boundary penetration to gripping methods suggests that the detailed loading state (e.g., fixed stress or fixed strain) may play an important role in liquid film penetration.  This is consistent with the discussion of the simulation results presented above.

In many experiments, Ga penetrates into Al bicrystals as a thin layer that lengthened at nearly a fixed rate,~\cite{Hugo:AlGaBicrystalTEM1999, Kozlova:AlGaSEM2005, Pereiro-Lopez:AlGaPRL2005, Ludwig:AlGaBicrystal2005, Pereiro-Lopez:AlGaInvasion2006} whereas in experiments in polycrystalline Al, the rate of propagation was more jerky and irregular.~\cite{Pereiro-Lopez:AlGaPolycrystal2004}  Commonly, the penetration rate is higher in polycrystals than in bicrystals presumably because of higher residual stresses and more degrees of freedom for the relative motion of individual grains.~\cite{Pereiro-Lopez:AlGaBiPoly2003}  The penetration rate was also observed to be very sensitive to grain boundary crystallography.  {\em In-situ} synchrotron observations of high energy $150\,^{\circ} /[110]$ tilt boundaries by Ludwig {\em et al.}~\cite{Ludwig:AlGaBicrystal2005} showed the Ga penetration rate was nearly constant with a velocity $v \approx 1$-$10~\mu$m/s.  {\em In-situ} TEM observations of several Al bicrystal foils by \citeauthor{Hugo:AlGaBicrystalTEM1999}~\cite{Hugo:AlGaBicrystalTEM1999} also showed a nearly constant Ga penetration rate, $v \approx 0.5$-$7~\mu$m/s, depending on the grain boundary crystallography.  These two experiments were performed under nominally similar conditions.  In our simulations, we found that the Ga penetration rate was also nearly constant, albeit with a velocity of 0.1-0.2 m/s which is significantly larger than experimentally observed in Al bicrystals.  The discrepancy is likely attributable to the fact that our simulation was performed at relatively high temperature ($T=600$ K) and with relatively large stresses (to overcome the time scale limitations of MD simulations).  The {\em in-situ} experiments were performed at room temperature with a stress of several MPa, while the simulations were performed at 600 K with a stress of at least 250 MPa.  Another possible reason for the faster penetration rates in the simulations as compared with the experiments is that the minimum observable Ga layer thickness in the simulations was $\sim 0.2$ nm versus at least $\sim 5$ nm in experiments.  

In the present simulations, we observed that the separation between the pair of grains that meet at a grain boundary (measured using markers that were some distance from the grain boundary) increased as the Ga-rich layer propagated down the grain boundary (Fig.~\ref{fig:GBseparation}).  A similar analysis was also performed in a series of experiments by Ludwig {\em et al.},~\cite{Ludwig:AlGaBicrystal2005,Ludwig:MSEA2000,Pereiro-Lopez:AlGaPRL2005} where the distance between two grains in Al bicrystals was measured by image correlation techniques.  They observed an increase of separation between the points by tens of nanometers during penetration even without an applied stress (but possibly subject to residual stresses).  Although the time and length scales are somewhat different, these observations agree  well with our simulation results.  The discrepancy between the magnitudes of the relative displacements (by two orders of magnitude) can be explained by the same stress and grain size effects described above.  Both the simulations (Fig.~\ref{fig:GBseparation}) and experiments~\cite{Pereiro-Lopez:AlGaPRL2005} show that the measured Ga layer thickness $w_{Ga}$ is nearly zero until a time well beyond the time at which significant grain displacement was observed to begin.  This suggests that the grain boundary opening below the liquid film tip is the results of the bond stretching associated with the stress field of the advancing Ga penetration front.

\citeauthor{Hugo:AlGaBicrystalTEM1999}~\cite{Hugo:AlGaBicrystalTEM1999} observed generation of a moving strain field at the Ga penetration front using TEM.  They proposed that the penetration front acts as a line defect with a singular strain field.  Our simulations also revealed that the moving Ga penetration front was accompanied by a dislocation with its own unique singular stress pattern (Fig.~\ref{fig:StressField}).  This dislocation seems to precede the advancing and thickening  Ga penetration layer.  It is interesting to note that in the absence of an applied strain, no dislocation forms [Fig.~\ref{fig:StressField}(a)] and the Ga penetration rate decreases with time (Fig.~\ref{fig:PenetrationRate}).  However, when a strain is applied, dislocations form and climb at a fixed rate [Fig.~\ref{fig:StressField}(b) and (c)] and the Ga penetration rate is time independent (Fig.~\ref{fig:PenetrationRate}).  This suggests that the constant Ga penetration rate observed in the strained solid is associated with the fixed rate of ``climb'' of dislocations.  Here, the applied strain plays essential roles in Ga penetration: to aid the nucleation of dislocations at the grain boundary and to keep the grain boundary open to allow fast Ga transport enough to move with the dislocations. 

Our simulations demonstrate that application of a stress significantly promotes liquid metal penetration, resulting in a change from a diffusive to fixed rate penetration mode.  The change of penetration kinetics is attributed to the nucleation and constant-rate climbing of grain boundary dislocations.  Why do the dislocations move down the boundary?  The dislocation sets up its own stress field; in the present geometry, it is compressive above the dislocation line and tensile below.  The chemical potential along the grain boundary is proportional to the grain-boundary traction~\cite{Rice:DiffusiveCrack} or $\sigma_{xx}$ and, hence, the chemical potential along the grain boundary changes abruptly at the dislocation.  Ga atoms in the grain boundary respond by jumping quickly from above the dislocation line to below it.  This, in turn, moves the dislocation down, yet preserves the stress discontinuity (i.e., the dislocation is intact).  This explains why the dislocation climbs down at a fixed rate.  How fast does the dislocation climb?  This can be determined by solving the coupled elasticity/diffusion problem.  Similar problems were addressed by Chuang~\cite{Rice:DiffusiveCrack, Chuang:DiffusiveCrack, Antipov:IntegroEqDiffusiveGrack} and Gao {\it et al.}~\cite{Gao:GBDwedge, Antipov:IntegroEqGBDwedge, Antipov:IntegroEquation} in the context of diffusive crack or wedge growth along a grain boundary subjected to an applied stress.~\cite{Antipov:IntegroEqDiffusiveGrack, Antipov:IntegroEqGBDwedge, Antipov:IntegroEquation}  The steady-state dislocation climb velocity $V$ in this model can be approximated as~\cite{Antipov:IntegroEqDiffusiveGrack} 
\begin{equation}
\label{eq:ClimbVelocity}
V \approx  \frac {\Omega D_{gb}} {k T} \frac {E b} {(1-\nu^2) l_{c}^2},  
\end{equation}
where $\Omega$ is the atomic volume of the species with grain boundary diffusivity $D_{gb}$, $kT$ is the thermal energy, $b$ is the Burgers vector, $E$ is the  Young's modulus, and $\nu$ is Poisson's ratio.  $l_{c}$ is a characteristic length associated with the jump in stress across the dislocation and should be of order of the dislocation core size (i.e., a few \AA) and can be found by solving the singular coupled elasticity/diffusion problem.~\cite{Antipov:IntegroEqDiffusiveGrack, Antipov:IntegroEqGBDwedge, Antipov:IntegroEquation}  Using values for Ga in Al in Eq.~(\ref{eq:ClimbVelocity}) yields $V \approx 0.1$ m/s, which is consistent with the dislocation climb velocity in the present simulations.  

The following model for the embrittlement of Al by Ga emerges.  First, Ga diffuses down the grain boundary in Al below the liquid groove root and, if the quantity of the inserted Ga is sufficiently large, a dislocation is nucleated at the grain boundary with the aid of the applied stress.  The dislocation establishes its own stress field; in the present geometry; it is compressive above the dislocation and tensile below it.  The first dislocation climbs down by stress-enhanced Ga hoping across the dislocation core, leaving a tail of Ga behind).  This Ga hopping leads to a constant dislocation climb rate that is independent of the remote applied stress.  Once the dislocation moves far enough from the groove root, another dislocation is nucleated.  It too climbs down the grain boundary at the same rate, resulting in a uniform spacing of climbing dislocations.  Each dislocation further relaxes the applied stress until it reaches a level too small for further dislocation nucleation.  At this point, however, the stress is not fully relieved. This residual stress is further relaxed by Ga diffusion down the boundary.  Since Ga weakens the Al bonds at the grain boundary,~\cite{Thomson:AlGaGB_Abinitio1, Thomson:AlGaGB_Abinitio2, Zhang:AlGaGB_Abinitio3, Sigle:AlGaTEMAPL} this Ga penetration leads to boundary decohesion, grain boundary opening, and crack formation/propagation (i.e., LME cracking).  This crack can be filled with liquid Ga.  The propagation of this ``crack'' leads to further Ga layer thickening.  This ``crack,'' Ga layer at the grain boundary and dislocations move down the grain boundary in unison.  The Ga penetration rate mirrors the dislocation climb rate and hence is time independent.

\section{\label{sec:level_7Conclusion} Conclusion}

Although LME exhibits a diverse set of fracture characteristics, depending on the solid-liquid metal couple, the penetration of nanometer-thick liquid metal films along the grain boundary is one of the hallmarks of the process that has been observed in the classical LME systems, such as Al-Ga, Cu-Bi and Ni-Bi.  We have employed EAM potentials optimized for Al-Ga binary alloys in the performance of a series of MD simulations of an Al bicrystal in contact with liquid Ga in the presence and absence of an applied stress.  Our simulations demonstrated how Ga penetrates along the $\Sigma 5~ 36.9 \,^{\circ} /[010]$ symmetric tilt boundary during the early stages of LME and how an applied stress enhances the Ga penetration.  The simulations capture many of the experimentally observed trends in Ga penetration of grain boundaries in Al.  The key atomistic mechanism at the tip of the advancing Ga penetration layer was identified through analysis of displacement fields and the stress distribution within the Al-Ga bicrystal system.  The interplay of stress and penetrating Ga atoms leads to the nucleation of a train of dislocations on the grain boundary below the liquid groove root which climbs down the grain boundary at a nearly constant rate.  The dislocation climb mechanism and the Ga penetration are coupled.  While the dislocations do relax part of the applied stress, the residual stresses keep the grain boundary open, thereby allowing more, fast Ga transport to the penetration front (i.e., Ga layer thickening process).  We believe that the coupled Ga transport and ``dislocation climb'' is the key to the anomalously fast, time-independent penetration of Ga along grain boundaries in Al.  The simulations explain most of the main features of a series of experimental studies of LME in Al-Ga over the past decade.



\begin{acknowledgments}
The authors gratefully acknowledge the support of Korea Science \& Engineering Foundation (H.-S. N) and the support of the US Department of Energy, Office of Fusion Energy Sciences, Grant No. DE-FG02-011ER54628 and its Computational Materials Science Network.
\end{acknowledgments}



\appendix




\end{document}